\documentclass[12pt]{amsart}

\usepackage[margin=1in]{geometry}
\usepackage{amsmath,amssymb,amsthm,textcomp}
\usepackage{amsfonts,graphicx}
\usepackage[mathscr]{eucal}
\usepackage{color}
\allowdisplaybreaks[1]
\theoremstyle{definition}

\numberwithin{equation}{section}

\newcommand{\ncom}{\newcommand}

\ncom{\beq}{\begin{equation}}
	\ncom{\eeq}{\end{equation}}
\ncom{\bea}{\begin{eqnarray*}}
	\ncom{\eea}{\end{eqnarray*}}
\ncom{\beqa}{\begin{eqnarray}}
	\ncom{\eeqa}{\end{eqnarray}}
\ncom{\nno}{\nonumber}
\ncom{\non}{\nonumber}
\ncom{\ds}{\displaystyle}
\ncom{\half}{\frac{1}{2}}
\ncom{\mbx}{\makebox{.25cm}}
\ncom{\hs}{\mbox{\hspace{.25cm}}}
\ncom{\rar}{\rightarrow}
\ncom{\Rar}{\Rightarrow}
\ncom{\noin}{\noindent}
\ncom{\bc}{\begin{center}}
	\ncom{\ec}{\end{center}}
\ncom{\sz}{\scriptsize}
\ncom{\rf}{\ref}
\ncom{\s}{\sqrt{2}}
\ncom{\sgm}{\sigma}
\ncom{\Sgm}{\Sigma}
\ncom{\psgm}{\sigma^{\prime}}
\ncom{\dt}{\delta}
\ncom{\Dt}{\Delta}
\ncom{\lmd}{\lambda}
\ncom{\Lmd}{\Lambda}
\ncom{\Th}{\Theta}
\ncom{\e}{\eta}
\ncom{\eps}{\epsilon}
\ncom{\pcc}{\stackrel{P}{>}}
\ncom{\lp}{\stackrel{L_{p}}{>}}
\ncom{\dist}{{\rm\,dist}}
\ncom{\sspan}{{\rm\,span}}
\ncom{\re}{{\rm Re\,}}
\ncom{\im}{{\rm Im\,}}
\ncom{\sgn}{{\rm sgn\,}}
\ncom{\ba}{\begin{array}}
	\ncom{\ea}{\end{array}}
\ncom{\hone}{\mbox{\hspace{1em}}}
\ncom{\htwo}{\mbox{\hspace{2em}}}
\ncom{\hthree}{\mbox{\hspace{3em}}}
\ncom{\hfour}{\mbox{\hspace{4em}}}
\ncom{\vone}{\vskip 2ex}
\ncom{\vtwo}{\vskip 4ex}
\ncom{\vonee}{\vskip 1.5ex}
\ncom{\vthree}{\vskip 6ex}
\ncom{\vfour}{\vspace*{8ex}}
\ncom{\norm}{\|\;\;\|}
\ncom{\integ}[4]{\int_{#1}^{#2}\,{#3}\,d{#4}}
\ncom{\vspan}[1]{{{\rm\,span}\{ #1 \}}}
\ncom{\dm}[1]{ {\displaystyle{#1} } }
\ncom{\ri}[1]{{#1} \index{#1}}

\newtheorem{remark}{\bf Remark}[section]

\newtheorem{definition}{Definition}[section]
\newtheoremstyle
{remarkstyle}
{}
{11pt}
{}
{}
{\bfseries}
{:}
{     }
{\thmname{#1} \thmnumber{#2} }

\theoremstyle{remarkstyle}

\def\eps{\varepsilon}

\begin{document}
	\title{\Large C\lowercase{losed-form estimates for missing counts in multidimensional incomplete tables}}

\author[Sayan Ghosh]{S. Ghosh}
\address{Department of Statistics,
 University of Haifa, Abba Khoushy Avenue 199, Haifa 3498838, Israel.}
 \email{sayan38@gmail.com}
\author{P. Vellaisamy}
\address{Department of Mathematics,
Indian Institute of Technology Bombay, Powai, Mumbai 400076, India.}
\email{pv@math.iitb.ac.in}

\begin{abstract}
A useful technique for analyzing incomplete tables is to model the missing data mechanisms of the variables using log-linear models. In this paper, we use log-linear parametrization and propose estimation methods for arbitrary three-way and $n$-dimensional incomplete tables. All possible cases in which data on one or more of the variables may be missing are considered. We provide simple closed form estimates of expected cell counts and parameters for the various missing data models. We also obtain explicit boundary estimates under nonignorable nonresponse models. Finally, a real-life dataset is analyzed to illustrate our results for modelling and estimation in multidimensional incomplete tables.
\end{abstract}

\maketitle

\section{Introduction}
Contingency tables are frequently used for the display and analysis of categorical data. Missing data in such tables pose a common problem in various epidemiological studies, clinical trials and social science studies. The results of analyses that improperly treat missing data can be biased and imprecise obscuring the underlying phenomena. So, the analysis of contingency tables with missing data, also called incomplete tables, is of practical interest. The two types of counts in such tables are (i) fully observed counts and (ii) partially classified margins (nonresponses). A systematic study of missing data involves three types of missingness mechanisms proposed in the literature (see Little and Rubin (2002)): missing completely at random (MCAR), missing at random (MAR) and not missing at random (NMAR). If the probability (of an observation being missing) is independent of both observed and unobserved data, then a mechanism is said to be MCAR. It is called MAR if conditional on the observed data, the probability is independent of unobserved data, and NMAR if the probability depends only on unobserved data. For likelihood inference, nonresponses are classified as either ignorable (when the missing data mechanism is MAR or MCAR) or nonignorable (when the missing data mechanism is NMAR). 

According to Little and Rubin (2002), an incomplete table may be analyzed using mainly the following techniques: complete case analysis (using only the fully observed counts), weighting, imputation and modelling. 
Various types of models for analyzing such incomplete tables are available in the literature, for example, the pattern-mixture model (see Little (1993), Park and Brown (1994), Forster and Smith (1998)) and the selection model (see Fay (1986), Baker and Laird (1988), Little and Rubin (2002)). Log-linear models have generally been used to study missing data mechanisms in incomplete tables (see Baker and Laird (1988), Baker, Rosenberger and Dersimonian (1992), Smith, Skinner and Clarke (1999), Clarke (2002), Clarke and Smith (2004, 2005)). Some of the estimation methods used are weighted least squares, maximum likelihood (ML) and Bayesian techniques.  

Baker, Rosenberger and Dersimonian (1992) used log-linear models for analyzing a two-way incomplete table with data missing in both variables, and obtained closed-form estimates of missing counts. In this paper, we extend the hierarchical log-linear parametrization to arbitrary three-dimensional and $n$-dimensional incomplete tables in general. We focus on log-linear models with main effects and two-way interactions among variables and their missing indicators. This is because higher order interactions are difficult to interpret and models with  such parameters may become non-identifiable. We consider all possible cases when data on one or more of the variables are missing in the above tables and derive explicit, closed-form formulae for estimates of expected cell counts under various missing data models. The formulae involve only observed cell counts or their sums, which simplifies the fitting of the models. 

Incomplete tables with data missing in at least one variable are common
in the social sciences and medical fields. For example, in the analysis of survey
data, the gender of each respondent is usually known. Suppose we are interested in the
association between two partially missing variables (say, income and education
level), stratified by gender. This is an example of a three-way incomplete table with data on two variables missing. While the EM algorithm  (see Dempster, Laird and Rubin (1977)) is available for such settings, it does not automatically produce asymptotic covariance matrices for the parameter estimates so that estimation of standard errors of the estimates becomes difficult. The rate of convergence of the EM algorithm also depends on the proportion of missing information for each parameter. So Meng and Rubin (1991) proposed a componentwise EM procedure, which is computationally expensive for covariance estimation. In this paper, we explicitly model the missing data mechanism of each variable which leads to a full likelihood specification and use ML estimation to obtain the parameter estimates. Unlike the EM algorithm, covariance estimates of the parameters can be calculated in the usual way by inverting the Fisher information while any of the common fit statistics can be used to compare the fits of different models. Besides estimating missing cell counts, we obtain closed-form estimates of joint, marginal and conditional probabilities of the variables and their missingness under various models. Also, estimates of the marginal odds ratios and their asymptotic variances are provided for each model. 

The problem of boundary solutions occurs in nonignorable models while using ML estimation. Such solutions occur when the MLE's of nonresponse cell probabilities are all zeros for certain levels of a variable, that is, they lie on the boundary of the parameter space. Some references to this problem for various incomplete tables include Baker, Rosenberger and Dersimonian (1992), Park, Kim and Kim (2014), Ghosh and Vellaisamy (Forthcoming). In this paper, we provide explicit closed-form MLE's of expected cell counts and other parameters if boundary solutions occur under nonignorable models for some three-way incomplete tables. 

The remaining part of the paper is organized as follows. In Section 2, we provide log-linear parametrizations and discuss estimation methods for three-way incomplete tables with data missing in one variable, two variables and all variables. We also discuss boundary solutions and their occurrence under NMAR models in each of the above tables. Section 3 extends the methodology and results in Section 2 to arbitrary $n$-dimensional incomplete tables. A real-life dataset is analyzed in Section 4 to illustrate the results in Section 2. Section 5 provides some concluding remarks.

\section{Log-linear parametrization for 3-dimensional incomplete tables}
For studying missing data mechanisms in an $I\times J\times 2\times 2$ incomplete table, Baker, Rosenberger and Dersimonian (1992) considered nine identifiable log-linear models. In this section, we use such hierarchical log-linear models (see Ghosh and Vellaisamy (2016)) for three-way contingency tables where data on at least one of the variables may be missing. Partially classified (supplementary) margins of a table are assumed to be positive. 

\subsection{Missing in one of the variables} 
Without loss of generality (WLOG), let data on $Y_{1}$ be missing. Denote the missing indicator for $Y_{1}$ by $R$, where $R = 1$ if $Y_{1}$ is observed and $R = 2$ otherwise. Then we have an $I\times J\times K\times 2$ table corresponding to  $Y_{1},~Y_{2},~Y_{3}$ and $R$ with cell counts ${\bf y} = \{y_{ijkx}\}$, where $1\leq i\leq I,~1\leq j\leq J,~1\leq k\leq K$ and $x = 1,2.$ Denote the vector of observed counts by ${\bf y_{\textrm{obs}}} = (\{y_{ijk1}\},\{y_{+jk2}\})$, where $\{y_{ijk1}\}$ are the fully observed counts, $\{y_{+jk2}\}$ are the supplementary margins and `+' means summation over levels of the corresponding variable. Let ${\bf\pi} = \{\pi_{ijkx}\}$ be the vector of cell probabilities, $\mu = \{\mu_{ijkx}\}$ be the vector of expected counts and $N = \sum_{i,j,k,x} y_{ijkx}$ be the total cell count. For $I=J=K=2$, the $2\times 2\times 2\times 2$ incomplete table is shown below. 
\vone
\begin{table}[ht]
	\caption{ $2\times 2\times 2\times 2$ Incomplete Table}\label{t1}
\begin{center}
$
\begin{array}{|c|c|c|cc|}\hline
& & & Y_{3} = 1 & Y_{3} = 2 \\ \hline
R = 1 & Y_{1} = 1 & Y_{2} = 1 & y_{1111} & y_{1121} \\   
& & Y_{2} = 2 & y_{1211} & y_{1221} \\ \hline 
& Y_{1} = 2 & Y_{2} = 1 & y_{2111} & y_{2121} \\   
& & Y_{2} = 2 & y_{2211} & y_{2221} \\ \hline 
R = 2 & \text{Missing} & Y_{2} = 1 & y_{+112} & y_{+122} \\   
& & Y_{2} = 2 & y_{+212} & y_{+222} \\ \hline 
\end{array}
$
\end{center}
\end{table}
\vone The log-linear model (with no three-way interactions) for this case is given by 
\begin{eqnarray}\label{eq11.1}
\log \mu_{ijkx} &=& \lambda + \lambda_{Y_{1}}(i) + \lambda_{Y_{2}}(j) + \lambda_{Y_{3}}(k) + \lambda_{R}(x) +\lambda_{Y_{1}Y_{2}}(i,j) + \lambda_{Y_{1}Y_{3}}(i,k) + \lambda_{Y_{2}Y_{3}}(j,k) \nonumber \\
& & + \lambda_{Y_{1}R}(i,x) + \lambda_{Y_{2}R}(j,x) + \lambda_{Y_{3}R}(k,x). 
\end{eqnarray}
Each log-linear parameter in (\ref{eq11.1}) satisfies the constraint that the sum over each of its arguments is 0, for example, $\sum_{i}\lambda_{Y_{1}Y_{3}}(i,k) = \sum_{k}\lambda_{Y_{1}Y_{3}}(i,k) = 0$. Define $a_{ijk} = \frac{P(R = 2\mid Y_{1} = i, Y_{2} = j, Y_{3} = k)}{P(R = 1\mid Y_{1} = i, Y_{2} = j, Y_{3} = k)} \\ = \frac{\pi_{ijk2}}{\pi_{ijk1}} = \frac{\mu_{ijk2}}{\mu_{ijk1}},$ which describes the missing data mechanism of $Y_{1}.$ It is the odds of $Y_{1}$ being missing. Then $\mu_{ijk2} = a_{ijk}\mu_{ijk1}.$ Also, $\sum_{i,j,k}\mu_{ijk1}(1 + a_{ijk}) = N$ and the joint probability $\pi_{ijk.} = \mu_{ijk1}(1 + a_{ijk})/N$, from which the marginals may be derived. Note that under (\ref{eq11.1}), $a_{ijk} = \exp[-2\{\lambda_R(1) + \lambda_{Y_{1}R}(i,1) + \lambda_{Y_{2}R}(j,1) + \lambda_{Y_{3}R}(k,1)\}]$. Denote $a_{ijk}$ by $\alpha_{i..}$ or $\alpha_{.j.}$ or $\alpha_{..k}$ or $\alpha_{...}$ if it depends on only $i$ or $j$ or $k$ or none of these, respectively. From Ghosh and Vellaisamy (2016), we have the following definition.
\begin{definition}\label{def1}
The missing mechanism of $Y_{1}$ under (\ref{eq11.1}) is NMAR if $a_{ijk} = \alpha_{i..}$, MAR if $a_{ijk} = \alpha_{.j.}$ or $\alpha_{..k}$ and MCAR if $a_{ijk} = \alpha_{...}$. 
\end{definition}
Under Poisson sampling for observed cell counts, the log-likelihood of ${\bf\mu}$ is 
\begin{equation}\label{eq2.1}
l({\bf\mu};{\bf y}_{\textrm{obs}}) = \sum_{i,j,k}y_{ijk1}\log \mu_{ijk1} + \sum_{j,k}y_{+jk2}\log \mu_{+jk2} - \sum_{i,j,k,x}\mu_{ijkx} + \Delta,
\end{equation} 
where $\Delta$ is some constant. The various missing data models and the MLE's under them are given as follows : \\
1. $\alpha_{i..}$ (NMAR for $Y_{1}$). \\
We have $\hat{\mu}_{ijk1} = y_{ijk1}$ and $\hat{\alpha}_{i..}$ satisfies $\sum_{i}\hat{\mu}_{ijk1}\hat{\alpha}_{i..} = y_{+jk2}$ $\forall~1\leq j\leq J, 1\leq k\leq K.$ \\
2. $\alpha_{.j.}$ (MAR for $Y_{1}$). \\
We have $\hat{\mu}_{ijk1} = \frac{y_{ijk1}y_{+jk+}y_{+j+1}}{y_{+jk1}y_{+j++}}$ and $\hat{\alpha}_{.j.} = \frac{y_{+j+2}}{y_{+j+1}}.$ \\
3. $\alpha_{..k}$ (MAR for $Y_{1}$). \\
We have $\hat{\mu}_{ijk1} = \frac{y_{ijk1}y_{+jk+}y_{++k1}}{y_{+jk1}y_{++k+}}$ and $\hat{\alpha}_{..k} = \frac{y_{++k2}}{y_{++k1}}.$ \\
4. $\alpha_{...}$ (MCAR for $Y_{1}$). \\
We have $\hat{\mu}_{ijk1} = \frac{y_{ijk1}y_{+jk+}y_{+++1}}{y_{+jk1}y_{++++}}$ and $\hat{\alpha}_{...} = \frac{y_{+++2}}{y_{+++1}}.$ 

From Ghosh and Vellaisamy (2016), boundary solutions occur if $\hat{\alpha}_{i..} \leq 0$ for at least one and at most $(I-1)$ values of $Y_{1}$. If any $\hat{\alpha}_{i..} < 0$, then boundary estimates are obtained by setting $\hat{\alpha}_{i..} = 0$ in (\ref{eq2.1}). For example, if $Y_{1} $ is binary with levels $1$ and $2$, and $\hat{\alpha}_{1..} = 0$ under Model 1, then the MLE's are
\begin{equation*}
\hat{\alpha}_{2..} = \frac{y_{+++2}}{y_{2++1}},~\hat{\mu}_{1jk1} = y_{1jk1},~\hat{\mu}_{2jk1} = \frac{(y_{2jk1} + y_{+jk2})y_{2++1}}{y_{+++2}}.
\end{equation*}
A perfect fit model is one in which the estimated expected counts are equal to the observed counts. Consider now the hypotheses $H_{0}$: the proposed model (among Models 1-4 mentioned above) fits the data, and $H_{1}$: the perfect fit model fits the data. Let $L_{0}$ and $L_{1}$ denote the maximized log-likelihood functions under the proposed and perfect fit models respectively. Then the likelihood ratio statistic for testing $H_{0}$ against $H_{1}$ is given by
\begin{eqnarray}\label{eq11.2}
G^{2} &=& -2(L_{0} - L_{1}) \nonumber \\ 
&=& -2\left[\sum_{i,j,k}y_{ijk1}\ln\left(\frac{\hat{\mu}_{ijk1}}{y_{ijk1}}\right) + \sum_{j,k}y_{+jk2}\ln\left(\frac{\sum_{i}\hat{\mu}_{ijk1}\hat{a}_{ijk}}{y_{+jk2}}\right) - \sum_{i,j,k}\hat{\mu}_{ijk1}(1 + \hat{a}_{ijk}) + N\right].
\end{eqnarray}
Note that $G^{2}$ follows $\chi^{2}_{\nu}$ asymptotically, where $\nu = (I + 1)JK$ (number of observed counts) $-$ number of free estimable parameters under the proposed model. If $Y_{1}$ is binary and boundary solutions occur under Model 1, then the boundary MLE's are obtained for the level of $Y_{1}$ corresponding to which $G^{2}$ is minimum.

\subsection{Missing in two of the variables}
WLOG, suppose data on $Y_{1}$ and $Y_{2}$ are missing. For $i=1,2,$ denote the missing indicator for $Y_{i}$ by $R_{i}$ such that $R_{i} = 1$ if $Y_{i}$ is observed and $R_{i} = 2$ otherwise. Then we have an $I\times J\times K\times 2\times 2$ table corresponding to $Y_{1},~Y_{2},~Y_{3},~R_{1}$ and $R_{2}$ with cell counts ${\bf y} = \{y_{ijkxs}\}$, where $1\leq i\leq I,~1\leq j\leq J,~1\leq k\leq K$ and $x,s = 1,2.$ Denote the vector of observed counts by ${\bf y_{\textrm{obs}}} = (\{y_{ijk11}\},\{y_{+jk21}\},\{y_{i+k12}\},\{y_{++k22}\}).$ Also, let ${\bf\pi} = \{\pi_{ijkxs}\}$ be the vector of cell probabilities, $\mu = \{\mu_{ijkxs}\}$ be the vector of expected counts and $N$ be the total cell count. For $I = J = K = 2$, the $2\times 2\times 2\times 2\times 2$ incomplete table is shown below. 
\vone\noindent  
\begin{table}[ht]
\caption{$2\times 2\times 2\times 2\times 2$ Incomplete Table}\label{t2}
\begin{center}
$
\begin{array}{|c|c|cc|cc|}\hline
& & & & Y_{3} = 1 & Y_{3} = 2 \\ \hline
R_{1} = 1 & Y_{1} = 1 & R_{2} = 1 & Y_{2} = 1 & y_{11111} & y_{11211} \\   
& & & Y_{2} = 2 & y_{12111} & y_{12211} \\ \hline 
& & R_{2} = 2 & \text{Missing} & y_{1+112} & y_{1+212} \\ \hline
& Y_{1} = 2 & R_{2} = 1 & Y_{2} = 1 & y_{21111} & y_{21211} \\   
& & & Y_{2} = 2 & y_{22111} & y_{22211} \\ \hline 
& & R_{2} = 2 & \text{Missing} & y_{2+112} & y_{2+212} \\ \hline
R_{1} = 2 & \text{Missing} & R_{2} = 1 & Y_{2} = 1 & y_{+1121} & y_{+1221} \\   
& & & Y_{2} = 2 & y_{+2121} & y_{+2221} \\ \hline 
& & R_{2} = 2 & \text{Missing} & y_{++122} & y_{++222} \\ \hline  
\end{array}
$
\end{center} 
\end{table}
\vone
The log-linear model (without three-way or higher order interactions) is given by 
\begin{eqnarray}\label{eq22.1}
\log \mu_{ijkxs} &=& \lambda + \lambda_{Y_{1}}(i) + \lambda_{Y_{2}}(j) + \lambda_{Y_{3}}(k) + \lambda_{R_{1}}(x) + \lambda_{R_{2}}(s) + \lambda_{Y_{1}Y_{2}}(i,j) + \lambda_{Y_{1}Y_{3}}(i,k)  \nonumber \\
& & + \lambda_{Y_{2}Y_{3}}(j,k) + \lambda_{Y_{1}R_{1}}(i,x) + \lambda_{Y_{2}R_{1}}(j,x) + \lambda_{Y_{3}R_{1}}(k,x) \nonumber \\
& & + \lambda_{Y_{1}R_{2}}(i,s) + \lambda_{Y_{2}R_{2}}(j,s) + \lambda_{Y_{3}R_{2}}(k,s) + \lambda_{R_{1}R_{2}}(x,s). 
\end{eqnarray}
Each log-linear parameter in (\ref{eq22.1}) satisfies the constraint that the sum over each of its arguments is 0. Define the following quantities 
\begin{eqnarray*}
a_{ijk} &=& \frac{P(R_{1}=2,R_{2} = 1\mid Y_{1}=i,Y_{2}=j,Y_{3}=k)}{P(R_{1}=1,R_{2} = 1\mid Y_{1}=i,Y_{2}=j,Y_{3}=k)} = \frac{\pi_{ijk21}}{\pi_{ijk11}} = \frac{\mu_{ijk21}}{\mu_{ijk11}}, \nonumber \\
b_{ijk} &=& \frac{P(R_{1}=1,R_{2} = 2\mid Y_{1}=i,Y_{2}=j,Y_{3}=k)}{P(R_{1}=1,R_{2}=1\mid Y_{1}=i,Y_{2}=j,Y_{3}=k)} = \frac{\pi_{ijk12}}{\pi_{ijk11}} = \frac{\mu_{ijk12}}{\mu_{ijk11}}. \nonumber
\end{eqnarray*}
Then the missing data mechanisms of $Y_{1}$ and $Y_{2}$ are described by $a_{ijk}$ and $b_{ijk}$, respectively. Note that $a_{ijk}$ is the conditional odds of $Y_{1}$ being missing given $Y_{2}$ is observed, while $b_{ijk}$ is the conditional odds of $Y_{2}$ being missing given $Y_{1}$ is observed. The odds ratio between $R_{1}$ and $R_{2}$ is
\begin{eqnarray*}
\theta &=& \frac{P(R_{1}=1,R_{2}=1\mid Y_{1}=i,Y_{2}=j,Y_{3}=k)P(R_{1}=2,R_{2}=2\mid Y_{1}=i,Y_{2}=j,Y_{3}=k)}{P(R_{1}=1,R_{2}=2\mid Y_{1}=i,Y_{2}=j,Y_{3}=k)P(R_{1}=2,R_{2}=1\mid Y_{1}=i,Y_{2}=j,Y_{3}=k)} \nonumber \\
&=& \frac{\pi_{ijk11}\pi_{ijk22}}{\pi_{ijk12}\pi_{ijk21}} = \frac{\mu_{ijk11}\mu_{ijk22}}{\mu_{ijk12}\mu_{ijk21}}. \nonumber
\end{eqnarray*} 
If $\theta = 1$, then the missingness patterns of $Y_{1}$ and $Y_{2}$, that is, $R_{1}$ and $R_{2}$ are independent. Also, $\mu_{ijk21} = a_{ijk}\mu_{ijk11}, \mu_{ijk12} = b_{ijk}\mu_{ijk11}$, $\mu_{ijk22} = \mu_{ijk11}a_{ijk}b_{ijk}\theta$ and $N = \sum_{i,j,k}\mu_{ijk11}(1 + a_{ijk} + b_{ijk} + a_{ijk}b_{ijk}\theta)$. The joint probability is $\pi_{ijk..} = \mu_{ijk11}(1 + a_{ijk} + b_{ijk} + a_{ijk}b_{ijk}\theta)/N$, from which the marginals can be obtained. The conditional probability of $Y_{1}$ being missing given that $Y_{2}$ is observed is 
\begin{equation*}
\phi_{1|2}(i,j,k) = P(R_{1}=2\mid R_{2}=1,Y_{1}=i,Y_{2}=j,Y_{3}=k) = \frac{a_{ijk}}{1+a_{ijk}}.
\end{equation*}
Similarly, the conditional probability of $Y_{2}$ being missing given that $Y_{1}$ is observed is 
\begin{equation*}
\phi_{2|1}(i,j,k) = P(R_{2}=2\mid R_{1}=1,Y_{1}=i,Y_{2}=j,Y_{3}=k) = \frac{b_{ijk}}{1+b_{ijk}}.
\end{equation*} 
Under (\ref{eq22.1}), we have $a_{ijk} = \exp[-2\{\lambda_{R_{1}}(1) + \lambda_{Y_{1}R_{1}}(i,1) + \lambda_{Y_{2}R_{1}}(j,1) + \lambda_{Y_{3}R_{1}}(k,1) + \lambda_{R_{1}R_{2}}(1,1)\}]$, $b_{ijk} = \exp[-2\{\lambda_{R_{2}}(1) + \lambda_{Y_{1}R_{2}}(i,1) + \lambda_{Y_{2}R_{2}}(j,1) + \lambda_{Y_{3}R_{2}}(k,1) + \lambda_{R_{1}R_{2}}(1,1)\}]$ and \\ $\theta = \exp[4\lambda_{R_{1}R_{2}}(1,1)]$. 
If each of $a_{ijk}$ and $b_{ijk}$ depends on only one of $i$, $j$, $k$ or none of these, then let $a_{ijk} \in\{\alpha_{i..},\alpha_{.j.},\alpha_{..k},\alpha_{...}\}$ and $b_{ijk} \in\{\beta_{i..},\beta_{.j.},\beta_{..k},\beta_{...}\}$. The next definition is due to Ghosh and Vellaisamy (2016). 
\begin{definition}\label{def2.2}
The missing mechanism of $Y_{1}$ under (\ref{eq22.1}) is NMAR if $a_{ijk} = \alpha_{i..}$, MAR if $a_{ijk} = \alpha_{.j.}$ or $\alpha_{..k}$ and MCAR if $a_{ijk} = \alpha_{...}$, respectively. Similarly, the missing mechanism of $Y_{2}$ is NMAR if $b_{ijk} = \beta_{.j.}$, MAR if $b_{ijk} = \beta_{i..}$ or $\beta_{..k}$ and MCAR if $b_{ijk} = \beta_{...}$. 
\end{definition}
\noindent Under Poisson sampling, the log-likelihood kernel of ${\bf\mu}$ is 
\begin{eqnarray}\label{eq2.2}
l({\bf\mu};{\bf y}_{\textrm{obs}}) &=& \sum_{i,j,k}y_{ijk11}\log \mu_{ijk11} + \sum_{j,k}y_{+jk21}\log \mu_{+jk21} + \sum_{i,k}y_{i+k12}\log \mu_{i+k12} \nonumber \\ 
& & + \sum_{k}y_{++k22}\log \mu_{++k22} - \sum_{i,j,k,x,s}\mu_{ijkxs}.
\end{eqnarray}
There are 16 identifiable models in this case. The various models and the MLE's under them are given in the Appendix. From Ghosh and Vellaisamy (2016), boundary solutions occur under at least one of the following cases. 
\begin{enumerate}
\item[1.] $\hat{\alpha}_{i..} \leq 0$ for at least one and at most $(I-1)$ values of $Y_{1}$, 
\item[2.] $\hat{\beta}_{.j.} \leq 0$ for at least one and at most $(J-1)$ values of $Y_{2}$. 
\end{enumerate}
They occur in models for which the missing mechanism of at least one of the variables is NMAR. If any $\hat{\alpha}_{i..} < 0$ or any $\hat{\beta}_{.j.} < 0$, then boundary estimates can still be obtained by setting $\hat{\alpha}_{i..} = 0$ or $\hat{\beta}_{.j.} = 0$ in (\ref{eq2.2}) for relevant models. 
Now suppose $Y_{1}$ and $Y_{2}$ are binary variables, each with levels $1$ and $2$. Then we have a $2\times 2\times K\times 2\times 2$ incomplete contingency table. The boundary MLE's obtained when $\hat{\alpha}_{1..} = 0$ or $\hat{\beta}_{.2.} = 0$ (say) under various NMAR models are shown below. \\
(a) $(\alpha_{i..},\beta_{...})$ (NMAR for $Y_{1}$, MCAR for $Y_{2}$) : \\
If $\hat{\alpha}_{1..} = 0$, then the MLE's are 
\begin{eqnarray*}
\hat{\alpha}_{2..} &=& \displaystyle\frac{y_{+++21}y_{+++1+}}{y_{+++11}y_{2++1+}},~\hat{\beta}_{...} = \displaystyle\frac{y_{+++12}}{y_{+++11}},~\hat{\theta} = \displaystyle\frac{y_{+++11}y_{+++22}}{y_{+++12}y_{+++21}},~\hat{\mu}_{1jk11} = \displaystyle\frac{y_{1jk11}y_{1++1+}y_{+++11}}{y_{1++11}y_{+++1+}}, \nonumber \\
\hat{\mu}_{2jk11} &=& \displaystyle\frac{y_{+++11}y_{2++1+}(y_{2jk11} + y_{+jk21})}{y_{+++1+}(y_{2++11} + y_{+++21})}. \nonumber
\end{eqnarray*}
(b) $(\alpha_{i..},\beta_{i..})$ (NMAR for $Y_{1}$, MAR for $Y_{2}$) : \\
If $\hat{\alpha}_{1..} = 0$, then the MLE's are 
\begin{eqnarray*}
\hat{\alpha}_{2..} &=& \displaystyle\frac{y_{+++21}}{y_{2++11}},~\hat{\beta}_{i..} = \displaystyle\frac{y_{i++12}}{y_{i++11}},~\hat{\theta} = \displaystyle\frac{y_{2++11}y_{+++22}}{y_{2++12}y_{+++21}}, \nonumber \\
\hat{\mu}_{1jk11} &=& y_{1jk11},~\hat{\mu}_{2jk11} = \displaystyle\frac{y_{2++11}(y_{2jk11} + y_{+jk21})}{y_{2++11} + y_{+++21}}.
\end{eqnarray*}
(c) $(\alpha_{i..},\beta_{.j.})$ (NMAR for both $Y_{1}$ and $Y_{2}$) : \\
(i) If $\hat{\alpha}_{1..} = 0$, then the MLE's are 
\begin{equation*}
\hat{\alpha}_{2..} = \displaystyle\frac{y_{+++21}}{y_{2++11}},~\hat{\theta} = \displaystyle\frac{y_{2++11}y_{+++22}}{y_{2++12}y_{+++21}},~\hat{\mu}_{1jk11} = y_{1jk11},~\hat{\mu}_{2jk11} = \displaystyle\frac{y_{2++11}(y_{2jk11} + y_{+jk21})}{y_{2++11} + y_{+++21}}. \nonumber
\end{equation*}
Also, $\hat{\beta}_{.j.}$ satisfies $\sum_{j}\hat{\mu}_{ijk11}\hat{\beta}_{.j.} = y_{i+k12}$. \\
(ii) If $\hat{\beta}_{.2.} = 0$, then the MLE's are
\begin{equation*}
\hat{\beta}_{.1.} = \displaystyle\frac{y_{+++12}}{y_{+1+11}},~\hat{\theta} = \displaystyle\frac{y_{+1+11}y_{+++22}}{y_{+1+12}y_{+++21}},~\hat{\mu}_{i1k11} = \displaystyle\frac{y_{+1+11}(y_{i1k11} + y_{i+k12})}{y_{+1+11} + y_{+++21}},~\hat{\mu}_{i2k11} =  y_{i2k11}. \nonumber
\end{equation*} 
Also, $\hat{\alpha}_{i..}$ satisfies $\sum_{i}\hat{\mu}_{ijk11}\hat{\alpha}_{i..} = y_{+jk21}$. \\
(d) $(\alpha_{...},\beta_{.j.})$ (NMAR for $Y_{2}$, MCAR for $Y_{1}$) : \\
If $\hat{\beta}_{.2.} = 0$, then the MLE's are
\begin{eqnarray*}
\hat{\beta}_{.1.} &=& \displaystyle\frac{y_{+++12}y_{++++1}}{y_{+++11}y_{+1++1}},~\hat{\alpha}_{...} = \displaystyle\frac{y_{+++12}}{y_{+++11}},~\hat{\theta} = \displaystyle\frac{y_{+++11}y_{+++22}}{y_{+++12}y_{+++21}},\nonumber \\
\hat{\mu}_{i1k11} &=& \displaystyle\frac{y_{+++11}y_{+1++1}(y_{i1k11} + y_{i+k12})}{y_{++++1}(y_{+1+11} + y_{+++12})},~ 
\hat{\mu}_{i2k11} = \displaystyle\frac{y_{i2k11}y_{+2++1}y_{+++11}}{y_{+2+11}y_{++++1}}. \nonumber
\end{eqnarray*}
(e) $(\alpha_{.j.},\beta_{.j.})$ (NMAR for $Y_{2}$, MAR for $Y_{1}$) : \\
If $\hat{\beta}_{.2.} = 0$, then the MLE's are
\begin{eqnarray*}
\hat{\beta}_{.1.} &=& \displaystyle\frac{y_{+++12}}{y_{+1+11}},~\hat{\alpha}_{.j.} = \displaystyle\frac{y_{+jk21}}{y_{+jk11}},~\hat{\theta} = \displaystyle\frac{y_{+1+11}y_{+++22}}{y_{+1+12}y_{+++21}},\nonumber \\
\hat{\mu}_{i1k11} &=& \displaystyle\frac{y_{+1+11}(y_{i1k11} + y_{i+k12})}{y_{+1+11} + y_{+++21}},~\hat{\mu}_{i2k11} = y_{i2k11}. 
\end{eqnarray*}
The above method for obtaining closed-form boundary MLE's can be generalized to non-binary variables also. Next consider testing the hypotheses $H_{0}$: the proposed model (among Models 1-16 in the Appendix) fits the data against $H_{1}$: the perfect fit model fits the data. Let $L_{0}$ and $L_{1}$ denote the maximized log-likelihood functions under the proposed and perfect fit models respectively. Then the likelihood ratio statistic for testing $H_{0}$ against $H_{1}$ is given by 
\begin{eqnarray}\label{eq22.2}
G^{2} &=& -2(L_{0} - L_{1}) \nonumber \\
&=& -2\left[\sum_{i,j,k}y_{ijk11}\ln\left(\frac{\hat{\mu}_{ijk11}}{y_{ijk11}}\right) + \sum_{j,k}y_{+jk21}\ln\left(\frac{\sum_{i}\hat{\mu}_{ijk11}\hat{a}_{ijk}}{y_{+jk21}}\right) \right. \nonumber \\  
& & \left. + \sum_{i,k}y_{i+k12}\ln\left(\frac{\sum_{j}\hat{\mu}_{ijk11}\hat{b}_{ijk}}{y_{i+k12}}\right) + \sum_{k}y_{++k22}\ln\left(\frac{\sum_{i,j}\hat{\mu}_{ijk11}\hat{a}_{ijk}\hat{b}_{ijk}\hat{\theta}}{y_{++k22}}\right) \right. \nonumber \\
& & \left. - \sum_{i,j,k}\hat{\mu}_{ijk11}(1 + \hat{a}_{ijk} + \hat{b}_{ijk} + \hat{a}_{ijk}\hat{b}_{ijk}\hat{\theta}) + N\right].
\end{eqnarray}
Note that $G^{2}$ follows $\chi^{2}_{\nu}$ asymptotically, where $\nu = (I + 1)(J + 1)K -$ number of free estimable parameters under the proposed model. If $Y_{1}$ and $Y_{2}$ are binary variables and boundary solutions occur, then the boundary MLE's are obtained for the level of $Y_{1}$ or $Y_{2}$ (depending on whether $\hat{\alpha}_{i..} < 0$ or $\hat{\beta}_{.j.} < 0$) corresponding to which $G^{2}$ is minimum. 

{\bf Marginal odds ratios}. 
When $Y_{3}=k$ is fixed, consider the $Y_{1}Y_{2}$-marginal odds ratios. Let $OR_{..k} = (\hat{\pi}_{ijk..}\hat{\pi}_{i'j'k..})/(\hat{\pi}_{ij'k..}\hat{\pi}_{i'jk..})$ denote an estimated odds ratio on the $Y_{1}Y_{2}$-margin, where $1\leq i < i'\leq I$, $1\leq j < j'\leq J$ and $1\leq k\leq K$. Also, let $OR_{11k} = (y_{ijk11}y_{i'j'k11})/(y_{ij'k11}y_{i'jk11})$ be the estimated odds ratio when $R_{1} = R_{2} = 1.$ From the closed-form MLE's for the models (see Appendix), it can be shown that $OR_{..k} = OR_{11k}$ under Models 2, 4, 9, 12, 13, 14 and 16 {\it a priori}, and under Models 1, 3, 5, 6, 8, 11 and 15 for non-boundary (interior) estimates. We can derive closed-form expressions for the asymptotic variance of estimated marginal odds ratio in case of non-boundary MLE's. We assume that the data follows Poisson distribution. The asymptotic variance of a statistic $f(\{y_{ijk11}\},\{y_{i+k12}\},\{y_{+jk21}\},y_{++k22})$ for fixed $k$ (see Baker (1994)) is
\begin{eqnarray}\label{eq22.3}
Var(f) &=& \sum_{i,j}\left(\frac{\partial f}{\partial y_{ijk11}}\right)^{2}\hat{\mu}_{ijk11} + \sum_{i}\left(\frac{\partial f}{\partial y_{i+k12}}\right)^{2}\hat{\mu}_{i+k12} + \sum_{j}\left(\frac{\partial f}{\partial y_{+jk21}}\right)^{2}\hat{\mu}_{+jk21} \nonumber \\
& & + \left(\frac{\partial f}{\partial y_{++k22}}\right)^{2}\hat{\mu}_{++k22}. 
\end{eqnarray}   
When $OR_{..k} = OR_{11k} = (y_{ijk11}y_{i'j'k11})/(y_{ij'k11}y_{i'jk11}),$, we get  from (\ref{eq22.3})
\begin{equation}\label{eq22.4}
Var(OR_{..k}) = OR_{..k}^{2}\left[\frac{\hat{\mu}_{ijk11}}{y^{2}_{ijk11}} + \frac{\hat{\mu}_{ij'k11}}{y^{2}_{ij'k11}} + \frac{\hat{\mu}_{i'jk11}}{y^{2}_{i'jk11}} + \frac{\hat{\mu}_{i'j'k11}}{y^{2}_{i'j'k11}}\right]. 
\end{equation}
Using (\ref{eq22.4}), the asymptotic variances of estimated marginal odds ratios for $k$ fixed under various models (see Appendix) are as follows. \\
1. Models 2, 3 and 4 : 
\begin{equation*}
Var(OR_{..k}) = OR_{..k}^{2}\frac{y_{++k11}}{y_{++k+1}}\left[\frac{y_{+jk+1}}{y_{+jk11}}\left(\frac{1}{y_{ijk11}} + \frac{1}{y_{i'jk11}}\right) + \frac{y_{+j'k+1}}{y_{+j'k11}}\left(\frac{1}{y_{ij'k11}} + \frac{1}{y_{i'j'k11}}\right)\right]    
\end{equation*}
2. Models 5, 9 and 13 : 
\begin{equation*}
Var(OR_{..k}) = OR_{..k}^{2}\frac{y_{++k11}}{y_{++k1+}}\left[\frac{y_{i+k1+}}{y_{i+k11}}\left(\frac{1}{y_{ijk11}} + \frac{1}{y_{ij'k11}}\right) + \frac{y_{i'+k1+}}{y_{i'+k11}}\left(\frac{1}{y_{i'jk11}} + \frac{1}{y_{i'j'k11}}\right)\right]
\end{equation*}
3. Models 6, 8, 11, 12, 14, 15 and 16 : 
\begin{equation*}
Var(OR_{..k}) = OR_{..k}^{2}\left[\frac{1}{y_{ijk11}} + \frac{1}{y_{ij'k11}} + \frac{1}{y_{i'jk11}} + \frac{1}{y_{i'j'k11}}\right]
\end{equation*}
It may be remarked that this variance approximation is based on a Taylor series linearization method (sometimes called the delta method). Alternatively, the variances can be computed from the inverse of the observed information matrix using the method in Baker (1991). Note that if boundary solutions occur under NMAR models, then this method provides a variance estimate given that the closed-form MLE's of the expected cell counts in (\ref{eq22.4}) lie on the boundary of the parameter space. However, the bootstrap technique provides an unconditional variance estimate in this case (see Baker and Laird (1988)). 

\subsection{Missing in all three variables}
For $i = 1,2,3,$ denote $R_{i}$ to be the missing indicator of $Y_{i}$, where $R_{i} = 1$ if $Y_{i}$ is observed and $R_{i} = 2$ otherwise. Then we have an $I\times J\times K\times 2\times 2\times 2$ table corresponding to $Y_{1},~Y_{2},~Y_{3},~R_{1},~R_{2}$ and $R_{3}$ with cell counts ${\bf y} = \{y_{ijkxsz}\}$, where $1\leq i\leq I,~1\leq j\leq J,~1\leq k\leq K$ and $x,s,z = 1,2.$ Also, ${\bf y_{\textrm{obs}}} = (\{y_{ijk111}\},\{y_{+jk211}\},\{y_{i+k121}\},\{y_{ij+112}\},\{y_{++k221}\},\{y_{+j+212}\},\{y_{i++122}\},y_{+++222})$. Let ${\bf\pi} = \{\pi_{ijkxsz}\}$ be the vector of cell probabilities, $N$ be the total cell count and $\mu = \{\mu_{ijkxsz}\}$ be the vector of expected counts. For $I = J= K = 2$, the $2\times 2\times 2\times 2\times 2\times 2$ incomplete table is shown below. \\

\begin{table}[ht]
\caption{$2\times 2\times 2\times 2\times 2\times 2$ Incomplete Table}\label{t3}
\begin{center}
$
\begin{array}{|c|c|cc|cc|c|}\hline
& & & & R_{3} = 1 & & R_{3} = 2 \\
& & & & Y_{3} = 1 & Y_{3} = 2 & \text{Missing} \\ \hline
R_{1} = 1 & Y_{1} = 1 & R_{2} = 1 & Y_{2} = 1 & y_{11111} & y_{112111} & y_{11+112} \\   
& & & Y_{2} = 2 & y_{121111} & y_{122111} & y_{12+112} \\ \hline 
& & R_{2} = 2 & \text{Missing} & y_{1+1121} & y_{1+2121} & y_{1++122} \\ \hline
& Y_{1} = 2 & R_{2} = 1 & Y_{2} = 1 & y_{211111} & y_{212111} & y_{21+112} \\   
& & & Y_{2} = 2 & y_{212111} & y_{222111} & y_{22+112} \\ \hline 
& & R_{2} = 2 & \text{Missing} & y_{2+1121} & y_{2+2121} & y_{2++122} \\ \hline
R_{1} = 2 & \text{Missing} & R_{2} = 1 & Y_{2} = 1 & y_{+11211} & y_{+12211} & y_{+1+212} \\   
& & & Y_{2} = 2 & y_{+21211} & y_{+22211} & y_{+2+212} \\ \hline 
& & R_{2} = 2 & \text{Missing} & y_{++1221} & y_{++2221} & y_{+++222} \\ \hline  
\end{array}
$
\end{center}
\end{table}
\noindent
The log-linear model in this case is 
\begin{eqnarray}\label{eq33.1}
\log \mu_{ijkxsz} &=& \lambda + \lambda_{Y_{1}}(i) + \lambda_{Y_{2}}(j) + \lambda_{Y_{3}}(k) + \lambda_{R_{1}}(x) + \lambda_{R_{2}}(s) + \lambda_{R_{3}}(z) + \lambda_{Y_{1}Y_{2}}(i,j) \nonumber \\
& & + \lambda_{Y_{1}Y_{3}}(i,k) + \lambda_{Y_{2}Y_{3}}(j,k) + \lambda_{Y_{1}R_{1}}(i,x) + \lambda_{Y_{2}R_{1}}(j,x) + \lambda_{Y_{3}R_{1}}(k,x) + \lambda_{Y_{1}R_{2}}(i,s) \nonumber \\& & + \lambda_{Y_{2}R_{2}}(j,s) + \lambda_{Y_{3}R_{2}}(k,s) + \lambda_{Y_{1}R_{3}}(i,z) +  \lambda_{Y_{2}R_{3}}(j,z) + \lambda_{Y_{3}R_{3}}(k,z) + \lambda_{R_{1}R_{2}}(x,s) \nonumber \\
& & + \lambda_{R_{1}R_{3}}(x,z) + \lambda_{R_{2}R_{3}}(s,z). 
\end{eqnarray}
Each log-linear parameter in (\ref{eq33.1}) satisfies the constraint that the sum over each of its arguments is 0. Define the following quantities 
\begin{eqnarray*}
a_{ijk} &=& \frac{P(R_{1}=2,R_{2}=1,R_{3}=1\mid Y_{1}=i,Y_{2}=j,Y_{3}=k)}{P(R_{1}=1,R_{2} = 1,R_{3}=1\mid Y_{1}=i,Y_{2}=j,Y_{3}=k)} = \frac{\pi_{ijk211}}{\pi_{ijk111}} = \frac{\mu_{ijk211}}{\mu_{ijk111}}, \nonumber \\
b_{ijk} &=& \frac{P(R_{1}=1,R_{2}=2,R_{3}=1\mid Y_{1}=i,Y_{2}=j,Y_{3}=k)}{P(R_{1}=1,R_{2}=1,R_{3}=1\mid Y_{1}=i,Y_{2}=j,Y_{3}=k)} = \frac{\pi_{ijk121}}{\pi_{ijk111}} = \frac{\mu_{ijk121}}{\mu_{ijk111}}, \nonumber \\
c_{ijk} &=& \frac{P(R_{1}=1,R_{2}=1,R_{3}=2\mid Y_{1}=i,Y_{2}=j,Y_{3}=k)}{P(R_{1}=1,R_{2}=1,R_{3}=1\mid Y_{1}=i,Y_{2}=j,Y_{3}=k)} = \frac{\pi_{ijk112}}{\pi_{ijk111}} = \frac{\mu_{ijk112}}{\mu_{ijk111}}. \nonumber
\end{eqnarray*}
Then $a_{ijk}$, $b_{ijk}$ and $c_{ijk}$ describe the missing data mechanisms of $Y_{1}$, $Y_{2}$ and $Y_{3}$, respectively. Here $a_{ijk}$ is the conditional odds of $Y_{1}$ being missing given both $Y_{2}$ and $Y_{3}$ are observed, $b_{ijk}$ is the conditional odds of $Y_{2}$ being missing given both $Y_{1}$ and $Y_{3}$ are observed, and $c_{ijk}$ is the conditional odds of $Y_{3}$ being missing given both $Y_{1}$ and $Y_{2}$ are observed. Let the conditional odds ratio between $R_{1}$ and $R_{2}$ given $Y_{3}$ is observed be
\begin{eqnarray*}
\theta_{12} &=& \frac{P(R_{1}=1,R_{2}=1,R_{3}=1|Y_{1}=i,Y_{2}=j,Y_{3}=k)}{P(R_{1}=1,R_{2}=2,R_{3}=1|Y_{1}=i,Y_{2}=j,Y_{3}=k)}\nonumber \\
&& \times\frac{P(R_{1}=2,R_{2}=2,R_{3}=1|Y_{1}=i,Y_{2}=j,Y_{3}=k)}{P(R_{1}=2,R_{2}=1,R_{3}=1|Y_{1}=i,Y_{2}=j,Y_{3}=k)} \nonumber \\
&=& \frac{\pi_{ijk111}\pi_{ijk221}}{\pi_{ijk121}\pi_{ijk211}} = \frac{\mu_{ijk111}\mu_{ijk221}}{\mu_{ijk121}\mu_{ijk211}}. \nonumber
\end{eqnarray*}
Similarly, define $\theta_{13}$ to be the conditional odds ratio between $R_{1}$ and $R_{3}$ given $Y_{2}$ is observed,
and $\theta_{23}$ to be the conditional odds ratio between $R_{2}$ and $R_{3}$ given $Y_{1}$ is observed.
Also, define
\begin{eqnarray*}
\theta_{123} &=& \frac{P(R_{1}=2,R_{2}=2,R_{3}=2\mid Y_{1}=i,Y_{2}=j,Y_{3}=k)}{P(R_{1}=1,R_{2}=1,R_{3}=1\mid Y_{1}=i,Y_{2}=j,Y_{3}=k)} \nonumber \\
&=& \frac{\pi_{ijk222}}{\pi_{ijk111}} = \frac{\mu_{ijk222}}{\mu_{ijk111}}.
\end{eqnarray*}
Here, $\theta_{12},\theta_{13}$ and $\theta_{23}$ describe the conditional associations between the missing mechanisms of $Y_{1}$ and $Y_{2}$, $Y_{1}$ and $Y_{3}$, and $Y_{2}$ and $Y_{3}$ respectively. For $i\neq j\neq k = 1,2,3,$ if $\theta_{ij} = 1,$ then the missing mechanisms of $Y_{i}$ and $Y_{j}$ are conditionally independent given that $Y_{k}$ is observed. Note that $\theta_{123}$ denotes the joint odds of $Y_{1},Y_{2}$ and $Y_{3}$ simultaneously missing. The joint probability is $\pi_{ijk..} = \mu_{ijk111}(1 + a_{ijk} + b_{ijk} + c_{ijk} + a_{ijk}b_{ijk}\theta_{12} + a_{ijk}c_{ijk}\theta_{13} + b_{ijk}c_{ijk}\theta_{23} + \theta_{123})/N$, from which the marginals can be obtained. Under (\ref{eq33.1}), we have
\begin{eqnarray*}
a_{ijk} &=& \exp[-2\{\lambda_{R_{1}}(1) + \lambda_{Y_{1}R_{1}}(i,1) + \lambda_{Y_{2}R_{1}}(j,1) + \lambda_{Y_{3}R_{1}}(k,1) + \lambda_{R_{1}R_{2}}(1,1) + \lambda_{R_{1}R_{3}}(1,1)\}], \nonumber \\
b_{ijk} &=& \exp[-2\{\lambda_{R_{2}}(1) + \lambda_{Y_{1}R_{2}}(i,1) + \lambda_{Y_{2}R_{2}}(j,1) + \lambda_{Y_{3}R_{2}}(k,1) + \lambda_{R_{1}R_{2}}(1,1) + \lambda_{R_{2}R_{3}}(1,1)\}], \nonumber \\
c_{ijk} &=& \exp[-2\{\lambda_{R_{3}}(1) + \lambda_{Y_{1}R_{3}}(i,1) + \lambda_{Y_{2}R_{3}}(j,1) + \lambda_{Y_{3}R_{3}}(k,1) + \lambda_{R_{1}R_{3}}(1,1) + \lambda_{R_{2}R_{3}}(1,1)\}], \nonumber \\
\theta_{12} &=& \exp[4\lambda_{R_{1}R_{2}}(1,1)],~\theta_{13} = \exp[4\lambda_{R_{1}R_{3}}(1,1)],~\theta_{23} = \exp[4\lambda_{R_{2}R_{3}}(1,1)], \nonumber \\
\theta_{123} &=& \exp[-2\{\lambda_{R_{1}}(1) + \lambda_{R_{2}}(1) + \lambda_{R_{3}}(1) + \lambda_{Y_{1}R_{1}}(i,1) + \lambda_{Y_{2}R_{1}}(j,1) + \lambda_{Y_{3}R_{1}}(k,1) \nonumber \\ 
& & + \lambda_{Y_{1}R_{2}}(i,1) + \lambda_{Y_{2}R_{2}}(j,1) + \lambda_{Y_{3}R_{2}}(k,1) + \lambda_{Y_{1}R_{3}}(i,1) + \lambda_{Y_{2}R_{3}}(j,1) + \lambda_{Y_{3}R_{3}}(k,1)\}]. \nonumber
\end{eqnarray*}
Based on the assumption in the previous case regarding the missing mechanism of a variable, $a_{ijk} \in\{\alpha_{...},\alpha_{i..},\alpha_{.j.},\alpha_{..k}\}$, $b_{ijk}\in\{ \beta_{...},\beta_{i..},\beta_{.j.},\beta_{..k}\}$ and $c_{ijk}\in\{ \gamma_{...},\gamma_{i..},\gamma_{.j.},\gamma_{..k}\}$ (say). For the definition below, see Ghosh and Vellaisamy (2016).
\begin{definition}\label{def3}
The missing mechanism of $Y_{1}$ under (\ref{eq33.1}) is NMAR if $a_{ijk} = \alpha_{i..}$, MAR if $a_{ijk} = \alpha_{.j.}$ or $\alpha_{..k}$ and MCAR if $a_{ijk} = \alpha_{...}$. Similarly, the missing mechanism of $Y_{2}$ is NMAR if $b_{ijk} = \beta_{.j.}$, MAR if $b_{ijk} = \beta_{i..}$ or $\beta_{..k}$ and MCAR if $b_{ijk} = \beta_{...}$. Finally, the missing mechanism of $Y_{3}$ is NMAR if $c_{ijk} = \gamma_{..k}$, MAR if $c_{ijk} = \gamma_{i..}$ or $\gamma_{.j.}$ and MCAR if $c_{ijk} = \gamma_{...}$.
\end{definition}
We have 64 possible identifiable models which are mixtures of the various missing mechanisms of the variables.
Under Poisson sampling, the log-likelihood can be wriiten as a function of  $a_{ijk},b_{ijk},c_{ijk},\theta_{12},\theta_{13},\theta_{23}$ and $\theta_{123}$ which is then maximized to obtain closed-form MLE's of $\mu_{ijkxsz}$ under various missing data models.
Note that from Ghosh and Vellaisamy (2016), boundary solutions occur if at least one of the following holds. 
\begin{enumerate}
\item[1.] $\hat{\alpha}_{i..} \leq 0$ for at least one and at most $(I-1)$ values of $Y_{1}$, 
\item[2.] $\hat{\beta}_{.j.} \leq 0$ for at least one and at most $(J-1)$ values of $Y_{2}$,
\item[3.] $\hat{\gamma}_{..k} \leq 0$ for at least one and at most $(K-1)$ values of $Y_{3}$. 
\end{enumerate}
The boundary estimates are obtained by setting $\hat{\alpha}_{i..} = 0$ or $\hat{\beta}_{.j.} = 0$ or $\hat{\gamma}_{..k} = 0$ in the log-likelihood for relevant models. The likelihood ratio statistic $G^{2}$ for testing the goodness of fit of a missing data model can be obtained as in the previous case. 
Here $G^{2}$ follows $\chi^{2}_{\nu}$ asymptotically, where $\nu = (I + 1)(J + 1)(K + 1) -$ number of free estimable parameters under the proposed model.
\begin{remark}\label{rem1}
For all the above cases, perfect fit solutions for fully observed counts occur under the following types of models: 
\begin{enumerate}
\item[(i)] non-boundary cases of NMAR only models for one or more variables, 
\item[(ii)] non-boundary cases of a mixture of NMAR and MAR models for the variables, 
\item[(iii)] MAR only models for two or more variables. 
\end{enumerate}
However, if the missing mechanism is MCAR for at least one of the variables, then perfect fit solutions don't occur.  
\end{remark}
WLOG, consider models in which the missing mechanism is NMAR for $Y_{1}$. Then we have the following observations.
\begin{remark}\label{rem2}
The systems of equations $\sum_{i}\hat{\mu}_{ijk1}\hat{\alpha}_{i..} = y_{+jk2},$ $\sum_{i}\hat{\mu}_{ijk11}\hat{\alpha}_{i..} = y_{+jk21}$ and $\sum_{i}\hat{\mu}_{ijk111}\hat{\alpha}_{i...} = y_{+jk211}$ for $I\times J\times K\times 2$, $I\times J\times K\times 2\times 2$ and $I\times J\times K\times 2\times 2\times 2$ tables respectively are overdetermined (underdetermined) if $I < JK~(I > JK)$. 
\end{remark}
\begin{remark}\label{rem3}
Let the matrix of coefficients be $A = (\hat{\mu}_{ijk1})$ or $A = (\hat{\mu}_{ijk11})$ or $A = (\hat{\mu}_{ijk111})$ for $I\times J\times K\times 2$ or $I\times J\times K\times 2\times 2$ or $I\times J\times K\times 2\times 2\times 2$ tables, respectively. \\
(a) Note that $A$ is square if $I = JK$ and rectangular otherwise from Remark \ref{rem2}. If $A$ is square and non-singular, then unique MLE's of $\alpha_{i..}$, $\beta_{.j.}$ and $\gamma_{..k}$ exist. \\
(b) For overdetermined systems in Remark \ref{rem2}, if $\text{rank}(A) = I$ (full rank), then the left inverse of $A$ exists and is given by $A^{-1}_{\text{left}} = (A^{T}A)^{-1}A^{T}$. Also, the unique solutions (MLE's of $\alpha_{i..}$, $\beta_{.j.}$ and $\gamma_{..k}$) are obtained using the method of ordinary least squares (see Williams (1990)). \\
(c) For underdetermined systems in Remark \ref{rem2}, if $\text{rank}(A) = JK$ (full rank), then the right inverse of $A$ exists and is given by $A^{-1}_{\text{right}} = A^{T}(AA^{T})^{-1}$. Also, the unique solutions (MLE's of $\alpha_{i..}$, $\beta_{.j.}$ and $\gamma_{..k}$) are obtained using the method of minimum norm least squares (see Madych (1991)).     
\end{remark}

\section{n-dimensional incomplete table}
In this section, we extend the discussions and results in the previous sections to $n$- dimensional incomplete tables.
\subsection{Log-linear parametrization}
Let $Y_{1},\ldots,Y_{n}$ be $n$ categorical variables with $I_{1},\ldots,I_{n}$ levels respectively. Assume data on $k$ of these variables are missing, while data on the remaining $(n-k)$ variables are always observed, where $1\leq k\leq n$. For $1\leq i\leq k,$ denote $R_{i}$ to be the missing indicator for $Y_{i}$, where $R_{i} = 1$ if data on $Y_{i}$ is observed and $R_{i} = 2$ otherwise. Accordingly, there are a variety of incomplete tables, from the $I_{1}\times I_{2}\times 2$ table (where one variable is missing) to the $I_{1}\times\ldots\times I_{n}\times 2^{n}$ table (where all $n$ variables are missing). There are $\binom{n}{k}$ ways in which data on $k$ variables may be missing. WLOG, we assume data on $Y_{1},\ldots,Y_{k}$ are missing. Then we have an $I_{1}\times\ldots\times I_{n}\times 2^{k}$ table. The vector of observed counts is 
\begin{eqnarray*}
{\bf y}_{\text{obs}} &=& (\{y_{i_{1}\ldots i_{n}1\ldots 1}\},\{y_{i_{1}+\ldots +i_{k+1}\ldots i_{n}12\ldots 21\ldots 1}\},\ldots,\{y_{+\ldots +i_{k}i_{k+1}\ldots i_{n}2\ldots 211\ldots 1}\},\ldots, \nonumber \\
& & \{y_{+\ldots +i_{k-1}i_{k}i_{k+1}\ldots i_{n}2\ldots 2111\ldots 1}\},\ldots,\{y_{i_{1}\ldots i_{k-1}+i_{k+1}\ldots i_{n}1\ldots 121\ldots 1}\},y_{+\ldots +2\ldots 2}). \nonumber 
\end{eqnarray*} 
Note that there are a total of $\prod_{k=1}^{n}I_{k}$ fully observed counts and $(2^{k}-1)$ supplementary margins. Let $\mu_{i_{1}\ldots i_{n}r_{1}\ldots r_{k}}= E(Y_{i_{1}\ldots i_{n}r_{1}\ldots r_{k}})$ denote the expected cell frequency. Then the log-linear model is given by
\begin{equation}\label{eq5.1}
\log \mu_{i_{1}\ldots i_{n}r_{1}\ldots r_{k}} = \lambda + \sum_{p=1}^{n}\lambda_{Y_{p}}(i_{p}) + \sum_{p\neq q=1}^{n}\lambda_{Y_{p}Y_{q}}(i_{p},i_{q}) + \sum_{p=1}^{n}\sum_{q=1}^{k}\lambda_{Y_{p}R_{q}}(i_{p},r_{q}) + \sum_{p\neq q=1}^{k}\lambda_{R_{p}R_{q}}(r_{p},r_{q}),
\end{equation}
where $1\leq i_{l}\leq I_{l},~1\leq l\leq n,~r_{j}=1,2,~1\leq j\leq k$. 

Three-way and higher order associations are assumed to be zero in (\ref{eq5.1}) as they are difficult to interpret. Also, closed-form MLE's of parameters become difficult to obtain along with issues of non-identifiability. Note that association terms among $Y_i$'s and those among $R_i$'s are not involved in studying the missing data mechanisms of $Y_i$'s in (\ref{eq5.1}). Hence, there is no need to include three-way or higher order interactions among the outcome variables such as $Y_{1}Y_{2}Y_{3}$ or the missing indicators such as $R_{1}R_{2}R_{3}$. It is assumed that the MAR mechanism of a variable depends on any one of the other variables so that interaction terms like $Y_{i}Y_{j}R_{k}$ for $i\neq j\neq k$ are excluded from (\ref{eq5.1}). The missingness mechanism of a variable cannot be NMAR and MAR simultaneously, which excludes terms with $Y_{i}Y_{j}R_{i}$ for $i \neq j$ in (\ref{eq5.1}). Interactions such as $Y_{i}R_{k}R_{l}$ for $i\neq k\neq l$ are absent in (\ref{eq5.1}) since their interpretation is unclear. Also, they are redundant for the derivation of closed-form estimates of the expected cell counts. The following constraints are required for identifiability of (\ref{eq5.1}) :
\begin{eqnarray*}
\sum_{i_{p}}\lambda_{Y_{p}}(i_{p}) &=& \sum_{i_{p}}\lambda_{Y_{p}Y_{q}}(i_{p},i_{q}) = \sum_{i_{q}}\lambda_{Y_{p}Y_{q}}(i_{p},i_{q}) = \sum_{i_{p}}\lambda_{Y_{p}R_{q}}(i_{p},r_{q}) = \sum_{r_{q}}\lambda_{Y_{p}R_{q}}(i_{p},r_{q}) \nonumber \\
&=& \sum_{r_{p}}\lambda_{R_{p}R_{q}}(r_{p},r_{q}) = \sum_{r_{q}}\lambda_{R_{p}R_{q}}(r_{p},r_{q}) = 0, \quad p\neq q.
\end{eqnarray*}
Next, we introduce some parameters to study the missingness mechanisms of $Y_{1},\ldots,Y_{k}.$
Define 
\begin{equation*}
\phi^{p}_{i_{1}\ldots i_{n}} = \frac{P(R_{1}=1,\ldots,R_{p}=2,\ldots,R_{k}=1\mid Y_{1}=i_{1},\ldots, Y_{n}=i_{n})}{P(R_{1}=1,\ldots,R_{p}=1,\ldots,R_{k}=1\mid Y_{1}=i_{1},\ldots, Y_{n}=i_{n})},\quad 1\leq p\leq k, 
\end{equation*}
which describes the missing data mechanism of $Y_{p}$. It is the conditional odds of $Y_{p}$ being missing given the other $Y_{i}$'s are observed. There are $k$ such odds. For $i\neq j\neq p=1,\ldots, k,$ define 
\begin{eqnarray*}
\theta_{ij} &=& \frac{P(R_{i}=1,R_{j}=1,\{R_{p}=1\}|Y_{1}=i_{1},\ldots,Y_{n}=i_{n})}{P(R_{i}=1,R_{j}=2,\{R_{p}=1\}|Y_{1}=i_{1},\ldots,Y_{n}=i_{n})}\nonumber \\
&& \times \frac{P(R_{i}=2,R_{j}=2,\{R_{p}=1\}|Y_{1}=i_{1},\ldots,Y_{n}=i_{n})}{P(R_{i}=2,R_{j}=1,\{R_{p}=1\}|Y_{1}=i_{1},\ldots,Y_{n}=i_{n})}, \nonumber
\end{eqnarray*}
which is the conditional odds ratio between $R_{i}$ and $R_{j}$. If $\theta_{ij} = 1,$ then the missingness patterns of $Y_{i}$ and $Y_{j}$, that is, $R_{i}$ and $R_{j}$  are conditionally independent given that the rest of $Y_{p}$'s are observed. There are $\binom{k}{2}$ such ratios. Let $A\subseteq \bar{k} = \{1,\ldots,k\}$ such that $|A|\geq 3.$ There are $(2^{k}-(k+1)-\binom{k}{2})$ such sets. Let $R_{A}=\{R_{i}|i\in A\}.$ Then $\{R_{A} = 1\} = \{R_{i}=1|i\in A\}$ and $\{R_{\bar{k}\backslash A} = 1\} = \{R_{i}=1|i\not\in A\}.$ Also, let $2_{A} = \{r_{i}=2|i\in A\}$, $1_{A} = \{r_{i}=1|i\in A\}$, $1_{\bar{k}\backslash A} = \{r_{i}=1|i\not\in A\}$, $2_{\bar{k}\backslash A} = \{r_{i}=2|i\not\in A\}$, $Y_{A} = \{Y_{i}|i\in A\}$ and $Y_{\bar{k}\backslash A} = \{Y_{i}|i\not\in A\}.$ Now define
\begin{equation*}
\theta_{A} = \frac{P(\{R_{A}=2\},\{R_{\bar{k}\backslash A}=1\}|Y_{1}=i_{1},\ldots,Y_{n}=i_{n})}{P(\{R_{A}=1\},\{R_{\bar{k}\backslash A}=1\}|Y_{1}=i_{1},\ldots,Y_{n}=i_{n})} = \frac{\pi_{i_{1}\ldots i_{n}2_{A}1_{\bar{k}\backslash A}}}{\pi_{i_{1}\ldots i_{n}1_{A}1_{\bar{k}\backslash A}}} = \frac{\mu_{i_{1}\ldots i_{n}2_{A}1_{\bar{k}\backslash A}}}{\mu_{i_{1}\ldots i_{n}1_{A}1_{\bar{k}\backslash A}}},
\end{equation*}  
which is the conditional odds of $Y_{A}$ being missing given that $Y_{\bar{k}\backslash A}$ are observed. Then for $1\leq p\leq k$ and $R_{p}=2,\{R_{\bar{k}\backslash \{p\}}=1\}$, we have $\mu_{i_{1}\ldots i_{n}1\ldots 2\ldots 1} = \phi^{p}_{i_{1}\ldots i_{n}}\mu_{i_{1}\ldots i_{n}1\ldots 1}.$ Also, $\mu_{i_{1}\ldots i_{n}1\ldots 1}\phi^{r}_{i_{1}\ldots i_{n}}\phi^{s}_{i_{1}\ldots i_{n}}\theta_{rs} = \mu_{i_{1}\ldots i_{n}2_{\{r,s\}}1_{\bar{k}\backslash \{r,s\}}}$ for $r\neq s=1,\ldots, k$ and $\mu_{i_{1}\ldots i_{n}1\ldots 1}\theta_{A} = \mu_{i_{1}\ldots i_{n}2_{A}1_{\bar{k}\backslash A}}$. Note that the joint probability 
\begin{equation*}
\pi_{i_{1}\ldots i_{n}+\ldots +} = \mu_{i_{1}\ldots i_{n}1\ldots 1}(1 + \sum_{p=1}^{k}\phi^{p}_{i_{1}\ldots i_{n}} \\ + \sum_{r\neq s =1}^{k}\phi^{r}_{i_{1}\ldots i_{n}}\phi^{s}_{i_{1}\ldots i_{n}}\theta_{rs} + \{\theta_{A}|A\subseteq\bar{k},|A|\geq 3\})/N,
\end{equation*}
from which the marginals can be obtained. The total count $N$ is obtained by summing both sides of the above equation over $i_{1},\ldots, i_{n}$. Under (\ref{eq5.1}), the parameters are given as follows.
\begin{eqnarray*}
\phi^{t}_{i_{1}\ldots i_{n}} &=& \exp\left[-2\left\{\lambda_{R_{t}}(1) + \sum_{p=1}^{n}\lambda_{Y_{p}R_{t}}(i_{p},1) + \sum_{p\neq t=1}^{k}\lambda_{R_{p}R_{t}}(1,1)\right\}\right],\quad 1\leq t\leq k, \\
\theta_{ij} &=& \exp\left[4\lambda_{R_{i}R_{j}}(1,1)\right],\quad i\neq j = 1,\ldots, k, \\
\theta_{A} &=& \exp\left[-2\left\{\sum_{p=1}^{k}\lambda_{R_{p}}(1) + \sum_{p=1}^{n}\sum_{q=1}^{k}\lambda_{Y_{p}R_{q}}(i_{p},1)\right\}\right], \quad A\subseteq \bar{k}, |A|\geq 3.  
\end{eqnarray*} 
The following definition (see Ghosh and Vellaisamy (2016)) gives the various missing data mechanisms of a variable under (\ref{eq5.1}). 
\begin{definition}\label{def4}
If $\phi^{p}_{i_{1}\ldots i_{n}}$ under (\ref{eq5.1}) depends on $i_{p}$ (denoted by $\phi^{p}_{\ldots i_{p}\ldots}$), then we have a NMAR missingness mechanism for $Y_{p}$. If it depends on $i_{q}$ for $p\neq q$ (denoted by $\phi^{p}_{\ldots i_{q}\ldots}$), then the missingness mechanism for $Y_{p}$ is MAR, while if it depends on none of $i_{1},\ldots,i_{n}$ (denoted by $\phi^{p}_{\ldots}$), then the missingness mechanism for $Y_{p}$ is MCAR. 
\end{definition}
\noindent
Since there are $(n+1)$ possible realizations of $\phi^{p}_{i_{1}\ldots i_{n}}$ for each $p=1,\ldots,k,$ we have a total of $(n+1)^{k}$ possible models which may be categorized as follows:
\begin{enumerate}
\item[B1.] MCAR model - the missingness mechanism of each of $Y_{1},\ldots,Y_{k}$ is constant (1 case), 
\item[B2.] NMAR model - the missingness mechanism of each of $Y_{1},\ldots,Y_{k}$ depends only on itself (1 case), 
\item[B3.] MAR model - the missingness mechanism of each of $Y_{1},\ldots,Y_{k}$ depends on any one of the remaining $(n-1)$ variables ($(n-1)^{k}$ cases), 
\item[B4.] Mixture of MCAR and NMAR models - the missingness mechanism of each of $Y_{1},\ldots,Y_{k}$ may be MCAR or NMAR, but all variables cannot have the same mechanism ($(2^{k}-2)$ cases), 
\item[B5.] Mixture of MCAR and MAR models - the missingness mechanism of each of $Y_{1},\ldots,Y_{k}$ may be MCAR or MAR, but all variables cannot have the same mechanism ($(n^{k} - (n-1)^{k} - 1)$ cases), 
\item[B6.] Mixture of NMAR and MAR models - the missingness mechanism of each of $Y_{1},\ldots,Y_{k}$ may be NMAR or MAR, but all variables cannot have the same mechanism ($(n^{k} - (n-1)^{k} - 1)$ cases), 
\item[B7.] Mixture of NMAR, MAR and MCAR models - the missingness mechanism of each of $Y_{1},\ldots,Y_{k}$ may be NMAR or MAR or MCAR, but all variables cannot have the same mechanism ($((n+1)^{k}+ (n-1)^{k} - 2(n^{k}-1) - 2^{k})$ cases).
\end{enumerate} 
The log-likelihood kernel under Poisson sampling is
\begin{eqnarray}\label{eq5.2}
l({\bf\mu};{\bf y}_{\textrm{obs}}) &=& \sum_{i_{1},\ldots,i_{n}}y_{i_{1}\ldots i_{n}1\ldots 1}\log \mu_{i_{1}\ldots i_{n}1\ldots 1} + \sum_{i_{2},\ldots,i_{n}}y_{+i_{2}\ldots i_{n}21\ldots 1}\log\mu_{+i_{2}\ldots i_{n}21\ldots 1} \nonumber \\
& & + \sum_{i_{1},\ldots, i_{k-1},i_{k+1},\ldots, i_{n}}y_{{i_{1}\ldots i_{k-1}+i_{k+1}\ldots i_{n}}1\ldots 121\ldots 1}\log\mu_{i_{1}\ldots i_{k-1}+i_{k+1}\ldots i_{n}1\ldots 121\ldots 1} + \ldots \nonumber \\
& & + \sum_{i_{k+1},\ldots,i_{n}}y_{+\ldots +i_{k+1}\ldots i_{n}2\ldots 21\ldots 1}\log\mu_{+\ldots +i_{k+1}\ldots i_{n}2\ldots 21\ldots 1} - \sum_{i_{1},\ldots,i_{n},r_{1},\ldots,r_{k}}\mu_{i_{1}\ldots i_{n}r_{1}\ldots r_{k}}. 
\end{eqnarray} 
Rewriting (\ref{eq5.2}) in terms of the parameters $\phi$'s and $\theta$'s, we can obtain closed-form MLE's of the parameters and the expected cell counts under the models described above. Perfect fits for fully observed counts are obtained for categories B2, B3 and B6 of models. From Ghosh and Vellaisamy (2016), boundary solutions occur if the MLE of any of the parameters $\phi$'s $< 0$, which are then set to zero to obtain boundary estimates. Note that for at least one $p\in\{1,\ldots,k\}$, we have $\hat{\phi}^{p}_{\ldots i_{p}\ldots} = 0$ for at least one and at most $(I_{p}-1)$ values of $Y_{p}$ in case of boundary solutions. 

Consider the hypotheses $H_{0}$: the proposed model (among models in categories B1 to B7 mentioned above) fits the data, and $H_{1}$: the perfect fit model fits the data. Let $L_{0}$ and $L_{1}$ denote the maximized log-likelihood functions under the proposed and perfect fit models respectively. Then the likelihood ratio statistic for testing $H_{0}$ against $H_{1}$ is 
\begin{eqnarray}\label{eq5.3}
G^{2} &=& -2(L_{0} - L_{1}) \nonumber \\
&=& -2\left[\sum_{i_{1},\ldots,i_{n}}\ln\left(\frac{\hat{\mu}_{i_{1}\ldots i_{n}1\ldots 1}}{y_{i_{1}\ldots i_{n}1\ldots 1}}\right) + \sum_{i_{2},\ldots,i_{n}}y_{+i_{2}\ldots i_{n}21\ldots 1}\ln\left(\frac{\sum_{i_{1}}\hat{\mu}_{i_{1}\ldots i_{n}1\ldots 1}\hat{\phi^{1}}_{i_{1}\ldots i_{n}}}{y_{+i_{2}\ldots i_{n}21\ldots 1}}\right) \right. \nonumber \\
& & \left. + \ldots + \sum_{i_{1},\ldots,i_{k-1},i_{k+1},\ldots,i_{n}}y_{i_{1}\ldots i_{k-1}+i_{k+1}\ldots i_{n}1\ldots 121\ldots 1}\ln\left(\frac{\sum_{i_{k}}\hat{\mu}_{i_{1}\ldots i_{n}1\ldots 1}\hat{\phi^{k}}_{i_{1}\ldots i_{n}}}{y_{i_{1}\ldots i_{k-1}+i_{k+1}\ldots i_{n}1\ldots 121\ldots 1}}\right) \right. \nonumber \\
& & \left. + \ldots + \sum_{i_{k+1},\ldots,i_{n}}y_{+\ldots +i_{k+1}\ldots i_{n}2\ldots 21\ldots 1}\ln\left(\frac{\hat{\mu}_{+\ldots +i_{k+1}\ldots i_{n}1\ldots 1}\hat{\theta}_{1\ldots k}}{y_{+\ldots +i_{k+1}\ldots i_{n}2\ldots 21\ldots 1}}\right) \right. \nonumber \\
& & \left. - \sum_{i_{1},\ldots i_{n}}\hat{\mu}_{i_{1}\ldots i_{n}1\ldots 1}\left(1 + \sum_{p=1}^{k}\hat{\phi}^{p}_{i_{1}\ldots i_{n}} + \sum_{r\neq s =1}^{k}\hat{\phi}^{r}_{i_{1}\ldots i_{n}}\hat{\phi}^{s}_{i_{1}\ldots i_{n}}\hat{\theta}_{rs} + \{\hat{\theta}_{A}|A\subseteq\bar{k},|A|\geq 3\}\right) + N\right].
\end{eqnarray}
Note that $G^{2}\sim\chi^{2}_{\nu}$ asymptotically, where $\nu = (\prod_{p=k+1}^{n}I_{p})\prod_{r\neq p=1}^{k}(1 + I_{r}) -$ number of free estimable parameters under the proposed model.

\section{Data Analysis}
In this section, we illustrate our results in Section 3 using a real-life example from Rubin, Stern and Vehovar (1995). Table \ref{t4} (a $2\times 2\times 2\times 2\times 2\times 2$ table) below shows the Slovenian public opinion (SPO) survey dataset classified by the variables Secession ($Y_{1}$), Attendance ($Y_{2}$) and Independence ($Y_{3}$), each having two levels Yes (1) and No (2). Here ``Missing" denotes the ``Don't know" category (missing margins) for each variable. Note that from (\ref{eq11.2}), (\ref{eq22.2}) and (\ref{eq5.3}), we observe that $G^{2}$ becomes undefined if any of the fully observed counts is $0$. So, the count 0 is replaced by 2 in the full table. 
\vone 
\begin{table}[ht]
\caption{Data from the SPO survey}\label{t4}
\begin{center}
$
\begin{array}{|c|c|cc|cc|c|}\hline
& & & & R_{3} = 1 & & R_{3} = 2 \\
& & & & Y_{3} = 1 & Y_{3} = 2 & \text{Missing} \\ \hline
R_{1} = 1 & Y_{1} = 1 & R_{2} = 1 & Y_{2} = 1 & 1191 & 8 & 21 \\   
& & & Y_{2} = 2 & 8 & 2 & 4 \\ \hline 
& & R_{2} = 2 & \text{Missing} & 107 & 3 & 9 \\ \hline
& Y_{1} = 2 & R_{2} = 1 & Y_{2} = 1 & 158 & 68 & 29 \\   
& & & Y_{2} = 2 & 7 & 14 & 3 \\ \hline 
& & R_{2} = 2 & \text{Missing} & 18 & 43 & 31 \\ \hline
R_{1} = 2 & \text{Missing} & R_{2} = 1 & Y_{2} = 1 & 90 & 2 & 109 \\   
& & & Y_{2} = 2 & 1 & 2 & 25 \\ \hline 
& & R_{2} = 2 & \text{Missing} & 19 & 8 & 96 \\ \hline  
\end{array}
$
\end{center}
\end{table}
\vone
WLOG, consider the subtable of Table \ref{t4} in which data on $Y_{1}$ is missing as shown below. 
\vone
\begin{table}[ht]
\caption{Subtable $Y_{1}$ of Table \ref{t4}}\label{t5}
\begin{center}
$
\begin{array}{|c|c|c|cc|}\hline
& & & Y_{3} = 1 & Y_{3} = 2 \\ \hline
R = 1 & Y_{1} = 1 & Y_{2} = 1 & 1191 & 8 \\   
& & Y_{2} = 2 & 8 & 2  \\ \hline 
& Y_{1} = 2 & Y_{2} = 1 & 158 & 68 \\   
& & Y_{2} = 2 & 7 & 14 \\ \hline
R = 2 & \text{Missing} & Y_{2} = 1 & 90 & 2 \\   
& & Y_{2} = 2 & 1 & 2 \\ \hline 
\end{array}
$
\end{center}
\end{table}
To determine the missing data mechanism, we fit Models 1-4 (see Section 2.1) to the data in Table \ref{t5}. The system of equations for Model 2 (NMAR for $Y_{1}$) yields $\hat{\alpha}_{1..} = 0.0721$ and $\hat{\alpha}_{2..} = 0.0258$ implying that boundary solutions do not occur. We use the closed-form MLE's in Section 2.1 to fit the above models. Let $G^{2}$ denote the likelihood ratio statistic for testing the goodness of fit of each of the Models 1-4 against the perfect fit model. The table below gives the $G^{2}$ values, $p$-values and degrees of freedom (d.f.) for the tests.
\vone
\begin{table}[ht]
\caption{Comparison of fit among models}\label{t6}
\begin{center}
$
\begin{array}{|c|c|c|c|c|}\hline
\textrm{Model} & \textrm{Boundary solution} & G^{2} & p\textrm{-value} & \textrm{d.f.} \\ \hline
\alpha_{i..} & \textrm{No} & 0 & 1 & 2 \\
\alpha_{.j.} & \textrm{No} & 2.4622 & 0.2920 & 2 \\
\alpha_{..k} & \textrm{No} & 2.0949 & 0.3508 & 2 \\ 
\alpha_{...} & \textrm{No} & 2.8538 & 0.4147 & 3 \\ \hline
\end{array}
$
\end{center}
\end{table}
\noindent
We usually don't consider perfect fit models (see the example in Baker, Rosenberger and Dersimonian (1992)) for model selection so that $\alpha_{i..}$ is discarded. From Table \ref{t6}, based on $p$-values, the plausible models for the data in Table \ref{t5} are $\alpha_{...}$, $\alpha_{.j.}$ and $\alpha_{..k}$. However, we deduce that the best fit model is $\alpha_{..k}$ (MAR for $Y_{1}$) based on minimum $G^{2}$ value = 2.0949. This implies that the missingness in the variable `Secession' depends on the observed variable `Independence'. This dependence is expected because if one is unsure about voting for Slovenian's secession from Yugoslavia, then one is also most likely decided about Slovenian independence. Note that `Secession' differs from `Independence' here since independence without secession was also possible with the formation of a new internal state.
\\
The table of expected cell counts using the closed-form estimates (see Section 2.1) is given below.
\vone
\begin{table}[ht]
\caption{Expected cell counts for model $\alpha_{..k}$ using closed-form estimates}\label{t7}
\begin{center}
$
\begin{array}{|c|c|c|cc|}\hline
& & & Y_{3} = 1 & Y_{3} = 2 \\ \hline
R = 1 & Y_{1} = 1 & Y_{2} = 1 & 1191.00 & 7.87 \\   
& & Y_{2} = 2 & 8.00 & 2.16 \\ \hline 
& Y_{1} = 2 & Y_{2} = 1 & 158.00 & 66.88 \\   
& & Y_{2} = 2 & 7.00 & 15.09 \\ \hline
R = 2 & Y_{1} = 1 & Y_{2} = 1 & 79.46 & 0.34 \\   
& & Y_{2} = 2 & 0.53 & 0.09 \\ \hline 
& Y_{1} = 2 & Y_{2} = 1 & 10.54 & 2.91 \\   
& & Y_{2} = 2 & 0.47 & 0.66 \\ \hline 
\end{array}
$
\end{center}
\end{table}
Next, consider WLOG the subtable of Table \ref{t4} in which data on $Y_{1}$ and $Y_{2}$ are missing as shown below. 
\vone
\begin{table}[ht]
\caption{ Subtable $Y_{1}Y_{2}$ of Table \ref{t4}}\label{t8}
\begin{center}
$
\begin{array}{|c|c|cc|cc|}\hline
& & & & Y_{3} = 1 & Y_{3} = 2 \\ \hline
R_{1} = 1 & Y_{1} = 1 & R_{2} = 1 & Y_{2} = 1 & 1191 & 8 \\   
& & & Y_{2} = 2 & 8 & 2 \\ \hline 
& & R_{2} = 2 & \text{Missing} & 107 & 3 \\ \hline
& Y_{1} = 2 & R_{2} = 1 & Y_{2} = 1 & 158 & 68 \\   
& & & Y_{2} = 2 & 7 & 14 \\ \hline 
& & R_{2} = 2 & \text{Missing} & 18 & 43 \\ \hline
R_{1} = 2 & \text{Missing} & R_{2} = 1 & Y_{2} = 1 & 90 & 2 \\   
& & & Y_{2} = 2 & 1 & 2 \\ \hline 
& & R_{2} = 2 & \text{Missing} & 19 & 8 \\ \hline  
\end{array}
$
\end{center}
\end{table}
To determine the missing data mechanism, we fit Models 1-16 (see Appendix) to the data in Table \ref{t8}. On solving the systems of equations in NMAR models for $Y_{1}$ or $Y_{2}$, we obtain $\hat{\alpha}_{1..} = 0.0721$, $\hat{\alpha}_{2..} = 0.0258$, $\hat{\beta}_{.1.} = 0.073$ and $\hat{\beta}_{.2.} = 2.375$. Hence, there are no boundary solutions. We use the closed-form MLE's in the Appendix to fit the above models . Let $G^{2}$ denote the likelihood ratio statistic for testing the goodness of fit of each of the Models 1-16 against the perfect fit model. The table below gives the $G^{2}$ values, $p$-values and degrees of freedom (d.f.) for the tests.
\vone
\begin{table}[ht]
\caption{Comparison of fit among models}\label{t9}
\begin{center}
$
\begin{array}{|c|c|c|c|c|}\hline
\textrm{Model} & \textrm{Boundary solution} & G^{2} & p\textrm{-value} & \textrm{d.f.} \\ \hline

(\alpha_{...},\beta_{i..}) & \textrm{No} & 48.1188 & < 0.0001 & 6 \\

(\alpha_{...},\beta_{.j.}) & \textrm{No} & 20.6256 & 0.0021 & 6 \\

(\alpha_{...},\beta_{..k}) & \textrm{No} & 4.5886 & 0.5975 & 6 \\

(\alpha_{i..},\beta_{...}) & \textrm{No} & 75.5003 & < 0.0001 & 6 \\

(\alpha_{i..},\beta_{i..}) & \textrm{No} & 49.7073 & < 0.0001 & 5 \\

(\alpha_{i..},\beta_{.j.}) & \textrm{No} & 14.7381 & 0.0115 & 5 \\

(\alpha_{i..},\beta_{..k}) & \textrm{No} & 2.8076 & 0.7296 & 5 \\

(\alpha_{.j.},\beta_{...}) & \textrm{No} & 75.1109 & < 0.0001 & 6 \\

(\alpha_{.j.},\beta_{i..}) & \textrm{No} & 45.9217 & < 0.0001 & 5 \\

(\alpha_{.j.},\beta_{.j.}) & \textrm{No} & 15.8222 & 0.0074 & 5 \\

(\alpha_{.j.},\beta_{..k}) & \textrm{No} & 4.2395 & 0.5155 & 5 \\

(\alpha_{..k},\beta_{...}) & \textrm{No} & 82.55 & < 0.0001 & 6 \\

(\alpha_{..k},\beta_{.i.}) & \textrm{No} & 50.6861 & < 0.0001 & 5 \\

(\alpha_{..k},\beta_{.j.}) & \textrm{No} & 17.8333 & 0.0032 & 5 \\

(\alpha_{..k},\beta_{..k}) & \textrm{No} & 5.4779 & 0.3604 & 5 \\ \hline

\end{array}
$
\end{center}
\end{table}
From Table \ref{t9}, based on $p$-values, the candidate models for the data in Table \ref{t8} are $(\alpha_{...},\beta_{..k})$, $(\alpha_{i..},\beta_{..k})$, $(\alpha_{.j.},\beta_{..k})$ and $(\alpha_{..k},\beta_{..k})$. However, we deduce that the best fit model is ($\alpha_{i..},\beta_{..k}$) (NMAR for $Y_{1}$, MAR for $Y_{2}$) based on minimum $G^{2}$ value = 2.8076. This implies that the missingness in the variable `Secession' depends on itself, while the missingness in `Attendance' depends on the variable `Independence'. This is due to the fact that if one is unsure about `Secession', then data on `Secession' will be missing. Also, if one is unsure about `Independence', then one may not attend the plebiscite. Hence, data on `Attendance' will be missing. 
The table of expected cell counts using the closed-form estimates (see Appendix) is given below.
\vone
\begin{table}[ht]
\caption{ Expected cell counts under model ($\alpha_{i..},\beta_{..k}$) using closed-form estimates}\label{t10}
\begin{center}
$
\begin{array}{|c|c|cc|cc|}\hline
& & & & Y_{3} = 1 & Y_{3} = 2 \\ \hline
R_{1} = 1 & Y_{1} = 1 & R_{2} = 1 & Y_{2} = 1 & 1191.00 & 8.00 \\   
& & & Y_{2} = 2 & 8.00 & 2.00 \\ \hline 
& & R_{2} = 2 & Y_{2} = 1 & 109.15 & 4.00 \\ 
& & & Y_{2} = 2 & 0.73 & 1.00 \\ \hline 
& Y_{1} = 2 & R_{2} = 1 & Y_{2} = 1 & 158.00 &  68.00 \\   
& & & Y_{2} = 2 & 7.00 & 14.00 \\ \hline 
& & R_{2} = 2 & Y_{2} = 1 & 14.48 & 34.00 \\ 
& & & Y_{2} = 2 & 0.64 & 7.00 \\ \hline 
R_{1} = 2 & Y_{1} = 1 & R_{2} = 1 & Y_{2} = 1 & 85.93 & 0.58 \\   
& & & Y_{2} = 2 & 0.58 & 0.14 \\ \hline 
& & R_{2} = 2 & Y_{2} = 1 & 21.84 & 0.80 \\ 
& & & Y_{2} = 2 & 0.15 & 0.20 \\ \hline 
& Y_{1} = 2 & R_{2} = 1 & Y_{2} = 1 & 4.07 & 1.75 \\   
& & & Y_{2} = 2 & 0.18 & 0.36 \\ \hline 
& & R_{2} = 2 & Y_{2} = 1 & 1.03 & 2.43 \\ 
& & & Y_{2} = 2 & 0.05 & 0.50 \\ \hline  
\end{array}
$
\end{center}
\end{table}
Note that $\hat\theta = 2.7738$ for the model ($\alpha_{i..},\beta_{..k}$), which implies that the missing mechanisms of the variables `Secession' and `Attendance' are probably not independent. That is, a realization is more likely to be missing for `Secession' if it is missing for `Attendance' or vice-versa. The estimated conditional probability of $Y_{1}$ being missing given $Y_{2}=1$ is observed is $\hat{\phi}_{1|2}(1) = \frac{\hat{\alpha}_{1..}}{1+\hat{\alpha}_{1..}}=0.0673$. Similarly, the estimated conditional probability of $Y_{1}$ being missing given $Y_{2}=2$ is observed is $\hat{\phi}_{1|2}(2) = \frac{\hat{\alpha}_{2..}}{1+\hat{\alpha}_{2..}}=0.0251$. So the estimated probability of nonresponse for `Secession' is greater when one replies `No' to attending the plebiscite. Also, the estimated conditional probability of $Y_{2}$ being missing given $Y_{1}=1$ is observed is $\hat{\phi}_{2|1}(1) = \frac{\hat{\beta}_{..1}}{1+\hat{\beta}_{..1}}=0.0839$. Similarly, the estimated conditional probability of $Y_{2}$ being missing given $Y_{1}=2$ is observed is $\hat{\phi}_{2|1}(2) = \frac{\hat{\beta}_{..2}}{1+\hat{\beta}_{..2}}=0.3333$. Hence, the estimated probability of nonresponse for `Attendance' is greater when one replies `No' to Slovenia's secession from Yugoslavia.

From the data in Table \ref{t8}, we have $OR_{..1} = OR_{111} = 6.5957$ and $OR_{..2} = OR_{112} = 0.8235$ for the model ($\alpha_{i..},\beta_{..k}$). This implies that if none of the responses for the variables is missing, then the estimated odds ratio between `Secession' and `Attendance' is greater when the response to `Independence' is `Yes' than when it is `No'. Also, $Var(OR_{..1}) = 11.9646$ and $Var(OR_{..2}) = 0.4823$, that is, for observed data, the estimated odds ratio between `Secession' and `Attendance' has greater precision when the response to `Independence' is `No' than when it is `Yes'.

\section{Conclusions}
In this paper, we have studied missing data mechanisms for variables in $I\times J\times K\times 2$, $I\times J\times K\times 2\times 2$ and $I\times J\times K\times 2\times 2\times 2$ incomplete contingency tables. For this purpose, we have considered hierarchical log-linear models which yield closed-form MLE's of parameters and expected cell counts under various missing data models. Closed-form estimates are also obtained for joint and marginal probabilities, marginal odds ratios, their asymptotic variances and conditional probabilities of missing variables under the models. Note that the methods and results in this paper are applicable for $I\times J\times 2$ and $I\times J\times 2\times 2$ tables also. Extensions of the models and estimation methods are presented for arbitrary $n$-dimensional incomplete tables. We have also provided closed-form boundary MLE's under various NMAR models in some incomplete tables. Finally, a real- life data analysis example validates our modelling approach and other results in this paper. 

 \section*{Appendix}
 The closed-form estimates of missing counts and other parameters under various missing data models for an $I\times J\times K\times 2\times 2$ table are as follows. \\
 \noindent
 1. $(\alpha_{...},\beta_{...})$ (MCAR for both $Y_{1}$ and $Y_{2}$). \\
 The MLE's are 
 \begin{equation*}
 \hat{\alpha}_{...} = \frac{y_{+++21}}{y_{+++11}},~\hat{\beta}_{...} = \frac{y_{+++12}}{y_{+++11}},~\hat{\theta} = \frac{y_{+++11}y_{+++22}}{y_{+++12}y_{+++21}}, 
 \end{equation*} 
 while the iterates of $\hat{\mu}_{ijk11}$ are
 \begin{equation*}
 \hat{\mu}^{(0)}_{ijk11} = y_{ijk11},~\hat{\mu}^{(t+1)}_{ijk11} = \frac{y_{+++11}\left(y_{ijk11}+\frac{y_{i+k12}}{\hat{\mu}^{(t)}_{i+k12}}.\hat{\mu}^{(t)}_{ijk11} + \frac{y_{+jk21}}{\hat{\mu}^{(t)}_{+jk21}}.\hat{\mu}^{(t)}_{ijk11}\right)}{y_{++++1}+y_{+++12}}. 
 \end{equation*}
 2. $(\alpha_{...},\beta_{i..})$ (MCAR for $Y_{1}$, MAR for $Y_{2}$). \\
 The MLE's are 
 \begin{equation*}
 \hat{\alpha}_{...} = \frac{y_{+++21}}{y_{+++11}},~\hat{\beta}_{i..} = \frac{y_{i++12}}{\hat{\mu}_{i++11}},~\hat{\theta} = \frac{y_{+++11}y_{+++22}}{y_{+++12}y_{+++21}},~\hat{\mu}_{ijk11} = \frac{y_{ijk11}y_{+++11}y_{+jk+1}}{y_{++++1}y_{+jk11}}.
 \end{equation*} 
 3. $(\alpha_{...},\beta_{.j.})$ (MCAR for $Y_{1}$, NMAR for $Y_{2}$). \\
 The MLE's are 
 \begin{equation*}
 \hat{\alpha}_{...} = \frac{y_{+++21}}{y_{+++11}},~\hat{\theta} = \frac{y_{+++11}y_{+++22}}{y_{+++12}y_{+++21}},~\hat{\mu}_{ijk11} = \frac{y_{ijk11}y_{+++11}y_{+jk+1}}{y_{++++1}y_{+jk11}}.
 \end{equation*}
 Also, $\hat{\beta}_{.j.}$ satisfies $\sum_{j}\hat{\mu}_{ijk11}\hat{\beta}_{.j.} = y_{i+k12}.$ \\
 4. $(\alpha_{...},\beta_{..k})$ (MCAR for $Y_{1}$, MAR for $Y_{2}$). \\
 The MLE's are 
 \begin{equation*}
 \hat{\alpha}_{...} = \frac{y_{+++21}}{y_{+++11}},~\hat{\beta}_{..k} = \frac{y_{++k12}}{\hat{\mu}_{++k11}},~\hat{\theta} = \frac{y_{+++11}y_{+++22}}{y_{+++12}y_{+++21}},~\hat{\mu}_{ijk11} = \frac{y_{ijk11}y_{+++11}y_{+jk+1}}{y_{++++1}y_{+jk11}}.
 \end{equation*}
 5. $(\alpha_{i..},\beta_{...})$ (NMAR for $Y_{1}$, MCAR for $Y_{2}$). \\
 The MLE's are
 \begin{equation*}
 \hat{\beta}_{...} = \frac{y_{+++12}}{y_{+++11}},~\hat{\theta} = \frac{y_{+++11}y_{+++22}}{y_{+++12}y_{+++21}},~\hat{\mu}_{ijk11} = \frac{y_{ijk11}y_{+++11}y_{i+k1+}}{y_{+++1+}y_{i+k11}}.
 \end{equation*}
 Also, $\hat{\alpha}_{i..}$ satisfies $\sum_{i}\hat{\mu}_{ijk11}\hat{\alpha}_{i..} = y_{+jk21}.$ \\
 6. $(\alpha_{i..},\beta_{i..})$ (NMAR for $Y_{1}$, MAR for $Y_{2}$). \\ 
 The MLE's are 
 \begin{equation*}
 \hat{\mu}_{ijk11} = y_{ijk11},~\hat{\beta}_{i..} = \frac{y_{i++12}}{y_{i++11}},~\hat{\theta} = \frac{y_{+++22}}{\sum_{i}y_{i++12}\hat{\alpha}_{i..}},
 \end{equation*}
 where $\hat{\alpha}_{i..}$ satisfies $\sum_{i}\hat{\mu}_{ijk11}\hat{\alpha}_{i..} = y_{+jk21}$. \\
 7. $(\alpha_{i..},\beta_{.j.})$ (NMAR for both $Y_{1}$ and $Y_{2}$). \\
 The MLE's are
 \begin{equation*}
 \hat{\mu}_{ijk11} = y_{ijk11},~\hat{\theta} = \frac{y_{+++22}}{\sum_{i,j}y_{ij+11}\hat{\alpha}_{i..}\hat{\beta}_{.j.}},
 \end{equation*}
 where $\hat{\alpha}_{i..}$ and $\hat{\beta}_{.j.}$ satisfy $\sum_{i}\hat{\mu}_{ijk11}\hat{\alpha}_{i..} = y_{+jk21}$ and $\sum_{j}\hat{\mu}_{ijk11}\hat{\beta}_{.j.} = y_{i+k12}$ respectively. \\
 8. $(\alpha_{i..},\beta_{..k})$ (NMAR for $Y_{1}$, MAR for $Y_{2}$). \\
 The MLE's are 
 \begin{equation*}
 \hat{\mu}_{ijk11} = y_{ijk11},~\hat{\beta}_{..k} = \frac{y_{++k12}}{y_{++k11}},~\hat{\theta} = \frac{y_{+++22}}{\sum_{i,k}y_{i+k11}\hat{\alpha}_{i..}\hat{\beta}_{..k}},
 \end{equation*}
 where $\hat{\alpha}_{i..}$ satisfies $\sum_{i}\hat{\mu}_{ijk11}\hat{\alpha}_{i..} = y_{+jk21}$. \\
 9. $(\alpha_{.j.},\beta_{...})$ (MAR for $Y_{1}$, MCAR for $Y_{2}$). \\ 
 The MLE's are
 \begin{equation*}
 \hat{\alpha}_{.j.} = \frac{y_{+j+21}}{\hat{\mu}_{+j+11}},~\hat{\beta}_{...} = \frac{y_{+++12}}{y_{+++11}},~\hat{\theta} = \frac{y_{+++11}y_{+++22}}{y_{+++12}y_{+++21}},~\hat{\mu}_{ijk11} = \frac{y_{ijk11}y_{+++11}y_{i+k1+}}{y_{+++1+}y_{i+k11}}.
 \end{equation*}
 10. $(\alpha_{.j.},\beta_{i..})$ (MAR for both $Y_{1}$ and $Y_{2}$). \\ 
 The MLE's are
 \begin{equation*}
 \hat{\mu}_{ijk11} = y_{ijk11},~\hat{\alpha}_{.j.} = \frac{y_{+j+21}}{y_{+j+11}},~\hat{\beta}_{i..} = \frac{y_{i++12}}{y_{i++11}},~\hat{\theta} = \frac{y_{+++22}}{\sum_{i,j}y_{ij+11}\hat{\alpha}_{.j.}\hat{\beta}_{i..}}.
 \end{equation*}
 11. $(\alpha_{.j.},\beta_{.j.})$ (MAR for $Y_{1}$, NMAR for $Y_{2}$). \\ 
 The MLE's are
 \begin{equation*}
 \hat{\mu}_{ijk11} = y_{ijk11},~\hat{\alpha}_{.j.} = \frac{y_{+j+21}}{y_{+j+11}},~\hat{\theta} = \frac{y_{+++22}}{\sum_{j}y_{+j+21}\hat{\beta}_{.j.}},
 \end{equation*} 
 where $\hat{\beta}_{.j.}$ satisfies $\sum_{j}\hat{\mu}_{ijk11}\hat{\beta}_{.j.} = y_{i+k12}$. \\
 12. $(\alpha_{.j.},\beta_{..k})$ (MAR for both $Y_{1}$ and $Y_{2}$). \\ 
 The MLE's are
 \begin{equation*}
 \hat{\mu}_{ijk11} = y_{ijk11},~\hat{\alpha}_{.j.} = \frac{y_{+j+21}}{y_{+j+11}},~\hat{\beta}_{..k} = \frac{y_{++k12}}{y_{++k11}},~\hat{\theta} = \frac{y_{+++22}}{\sum_{j,k}y_{+jk11}\hat{\alpha}_{.j.}\hat{\beta}_{..k}}.
 \end{equation*}
 13. $(\alpha_{..k},\beta_{...})$ (MAR for $Y_{1}$, MCAR for $Y_{2}$). \\ 
 The MLE's are
 \begin{equation*}
 \hat{\alpha}_{..k} = \frac{y_{++k21}}{\hat{\mu}_{++k11}},~\hat{\beta}_{...} = \frac{y_{+++12}}{y_{+++11}},~\hat{\theta} = \frac{y_{+++11}y_{+++22}}{y_{+++12}y_{+++21}},~\hat{\mu}_{ijk11} = \frac{y_{ijk11}y_{+++11}y_{i+k1+}}{y_{+++1+}y_{i+k11}}.
 \end{equation*}
 14. $(\alpha_{..k},\beta_{i..})$ (MAR for both $Y_{1}$ and $Y_{2}$). \\ 
 The MLE's are
 \begin{equation*}
 \hat{\mu}_{ijk11} = y_{ijk11},~\hat{\alpha}_{..k} = \frac{y_{++k21}}{y_{++k11}},~\hat{\beta}_{i..} = \frac{y_{i++12}}{y_{i++11}},~\hat{\theta} = \frac{y_{+++22}}{\sum_{i,k}y_{i+k11}\hat{\alpha}_{..k}\hat{\beta}_{i..}}.
 \end{equation*}
 15. $(\alpha_{..k},\beta_{.j.})$ (MAR for $Y_{1}$, NMAR for $Y_{2}$). \\ 
 The MLE's are
 \begin{equation*}
 \hat{\mu}_{ijk11} = y_{ijk11},~\hat{\alpha}_{..k} = \frac{y_{++k21}}{y_{++k11}},~\hat{\theta} = \frac{y_{+++22}}{\sum_{j,k}y_{+jk11}\hat{\alpha}_{..k}\hat{\beta}_{.j.}},
 \end{equation*} 
 where $\hat{\beta}_{.j.}$ satisfies $\sum_{j}\hat{\mu}_{ijk11}\hat{\beta}_{.j.} = y_{i+k12}$. \\
 16. $(\alpha_{..k},\beta_{..k})$ (MAR for both $Y_{1}$ and $Y_{2}$). \\ 
 The MLE's are
 \begin{equation*}
 \hat{\mu}_{ijk11} = y_{ijk11},~\hat{\alpha}_{..k} = \frac{y_{++k21}}{y_{++k11}},~\hat{\beta}_{..k} = \frac{y_{++k12}}{y_{++k11}},~\hat{\theta} = \frac{y_{+++22}}{\sum_{k}y_{++k12}\hat{\alpha}_{..k}}.
 \end{equation*} 
 Note that closed-form MLE's of $m_{jk11}$ exist for all models except for Model 1. In this case, $\hat{\mu}_{ijk11}$ may be obtained using the EM algorithm (see Dempster, Laird and Rubin (1977)).

\end{document}